\documentclass[prl,aps,twocolumn,preprintnumbers, showpacs, nofootinbib,superscriptaddress,notitlepage]{revtex4-1}
\usepackage{amssymb, amsthm}
\usepackage{amsmath}    
\usepackage{mathrsfs}   
\usepackage{graphicx}   
\usepackage{color}      
\usepackage{colortbl}
\usepackage{footnote}   
\usepackage{slashed}    

\usepackage[utf8]{inputenc}
\usepackage[dvipsnames]{xcolor}
\usepackage[normalem]{ulem}

\usepackage{subfiles}

\newcommand{\nn}{\nonumber}

\begin{document}


\title{Nucleon Transversity Distribution in the Continuum and Physical Mass Limit\\ from Lattice QCD}

\collaboration{\bf{Lattice Parton Collaboration ($\rm {\bf LPC}$)}}

{}
\affiliation{School of Physics and Electronics, Central South University, Changsha 418003, China}

\author{\includegraphics[scale=0.1]{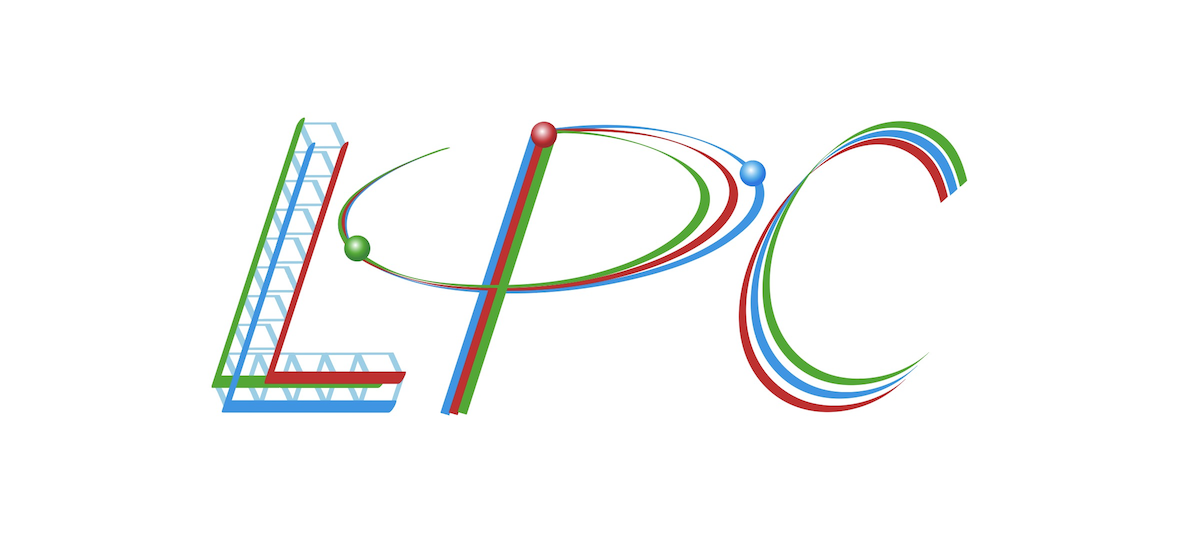}\\Fei Yao}
\affiliation{Center of Advanced Quantum Studies, Department of Physics, Beijing Normal University, Beijing 100875, China}

\author{Lisa Walter}
\affiliation{Institut f\"ur Theoretische Physik, Universit\"at Regensburg, D-93040 Regensburg, Germany}

\author{Jiunn-Wei Chen}
\affiliation{Department of Physics, Center for Theoretical Physics, and Leung Center for Cosmology and Particle Astrophysics, National Taiwan University, Taipei 10617, Taiwan}
\affiliation{Physics Division, National Center for Theoretical Sciences, Taipei 10617, Taiwan}

\author{Jun Hua}
\affiliation{Guangdong Provincial Key Laboratory of Nuclear Science, Institute of Quantum Matter, South China Normal University, Guangzhou 510006, China}
\affiliation{Guangdong-Hong Kong Joint Laboratory of Quantum Matter, Southern Nuclear Science Computing Center, South China Normal University, Guangzhou 510006, China}

\author{Xiangdong Ji}
\affiliation{Department of Physics, University of Maryland, College Park, MD 20742, USA}

\author{Luchang Jin}
\affiliation{Physics Department, University of Connecticut,
Storrs, Connecticut 06269-3046, USA}
\affiliation{RIKEN BNL Research Center, Brookhaven National Laboratory,
Upton, NY 11973, USA}

\author{Sebastian Lahrtz}
\affiliation{Institut f\"ur Theoretische Physik, Universit\"at Regensburg, D-93040 Regensburg, Germany}

\author{Lingquan Ma}
\affiliation{Center of Advanced Quantum Studies, Department of Physics, Beijing Normal University, Beijing 100875, China}

\author{Protick Mohanta}
\affiliation{Center of Advanced Quantum Studies, Department of Physics, Beijing Normal University, Beijing 100875, China}
\affiliation{School of Physical Sciences, National Institute of Science
Education and Research, HBNI, Odisha 752050, India}

\author{Andreas Sch\"afer}
\affiliation{Institut f\"ur Theoretische Physik, Universit\"at Regensburg, D-93040 Regensburg, Germany}

\author{Hai-Tao Shu}
\affiliation{Institut f\"ur Theoretische Physik, Universit\"at Regensburg, D-93040 Regensburg, Germany}


\author{Yushan Su}
\affiliation{Department of Physics, University of Maryland, College Park, MD 20742, USA}

\author{Peng Sun}
\email{Corresponding author: pengsun@impcas.ac.cn}
\affiliation{Institute of Modern Physics, Chinese Academy of Sciences, Lanzhou, Gansu Province 730000, China}
\affiliation{Department of Physics and Institute of Theoretical Physics,
Nanjing Normal University, Nanjing, Jiangsu 210023, China}

\author{Xiaonu Xiong}
\email{Corresponding author: xnxiong@csu.edu.cn}
\affiliation{School of Physics and Electronics, Central South University, Changsha 418003, China}

\author{Yi-Bo Yang}
\affiliation{CAS Key Laboratory of Theoretical Physics, Institute of Theoretical Physics, Chinese Academy of Sciences, Beijing 100190, China}
\affiliation{School of Fundamental Physics and Mathematical Sciences, Hangzhou Institute for Advanced Study, UCAS, Hangzhou 310024, China}
\affiliation{International Centre for Theoretical Physics Asia-Pacific, Beijing/Hangzhou, China}
\affiliation{University of Chinese Academy of Sciences, School of Physical Sciences, Beijing 100049, China}

\author{Jian-Hui Zhang}
\affiliation{Center of Advanced Quantum Studies, Department of Physics, Beijing Normal University, Beijing 100875, China}



\begin{abstract}
We report a state-of-the-art lattice QCD calculation of the isovector quark transversity distribution of the proton in the continuum and physical mass limit using large-momentum effective theory. The calculation is done at four lattice spacings $a=\{0.098,0.085,0.064,0.049\}$~fm and various pion masses ranging between $220$ and $350$ MeV, with proton momenta up to $2.8$ GeV. The result is non-perturbatively renormalized in the hybrid scheme with self renormalization which treats the infrared physics at large correlation distance properly, and extrapolated to the continuum, physical mass and infinite momentum limit. We also compare with recent global analyses for the nucleon isovector quark transversity distribution.
\end{abstract}

\maketitle

{\em Introduction:}
Parton distribution functions (PDFs) characterize the internal structure of hadrons in terms of the number densities of their quark and gluon constituents. They are crucial inputs for interpreting the experimental data collected at high-energy colliders such as the LHC. At leading-twist accuracy, there exist three quark PDFs: the unpolarized, the helicity and the transversity PDF~\cite{Ralston:1979ys}. Among them, the transversity PDF describes the correlation between the transverse polarizations of the nucleon and its quark constituents, and thus play an important role in describing the transverse spin structure of the nucleon~\cite{Jaffe:1991kp}. In contrast to the unpolarized and helicity PDFs, the transversity PDF is much less constrained from experiments. This is because it is a chiral-odd quantity, and has to couple to another chiral-odd quantity, such as the chiral-odd fragmentation or distribution function, in order to be measurable experimentally~\cite{Jaffe:1991ra,Cortes:1991ja,Jaffe:1993xb}. Currently, our knowledge of the transversity PDF mainly comes from measuring certain spin asymmetries in semi-inclusive deep-inelastic scattering or electron-positron annihilation to a hadron pair and Drell-Yan processes~\cite{Gamberg:2022kdb,Constantinou:2020hdm}. Based on these data, various global analyses of the transversity PDF have been performed~\cite{Anselmino:2007fs,Anselmino:2008jk,Anselmino:2013vqa,Kang:2014zza,Kang:2015msa,Bacchetta:2011ip,Bacchetta:2012ty,Radici:2015mwa,Lin:2017stx,Radici:2018iag,Cammarota:2020qcw,Gamberg:2022kdb}. 
Such analyses are expected to be greatly improved when more accurate data are accumulated at ongoing and future experiments at, e.g., the JLab 12 GeV upgrade and the Electron-Ion Collider (EIC).

On the other hand, recent theoretical developments~\cite{Braun:2007wv,Ji:2013dva,Ji:2014gla,Ma:2017pxb,Lin:2017snn,Radyushkin:2017cyf} have allowed us to calculate/fit the Bjorken $x$-dependence of PDFs from first-principle lattice QCD. Such calculations are particularly important for quantities like the nucleon transversity PDF which are hard to extract from experiments. In the past few years, several lattice calculations~\cite{Alexandrou:2018eet,Liu:2018hxv,HadStruc:2021qdf} of the nucleon transversity PDF have been carried out using either the large-momentum effective theory (LaMET)~\cite{Ji:2013dva,Ji:2014gla,Ji:2020ect} or the short-distance expansion (pseudo-PDF)~\cite{Radyushkin:2017cyf}. However, they were all done at a single lattice spacing, while a reliable extrapolation to the continuum is required to make a comparison with experimental measurements. Moreover, the non-perturbative renormalization used in these calculations has been shown to suffer from undesired infrared (IR) effects arising from the improper renormalization at long distances~\cite{Ji:2020brr}. Therefore, it is highly desirable to obtain a reliable prediction that uses 
a proper renormalization and is valid in the continuum and physical mass limit. This is the purpose of the present work. 

In this work, we present a state-of-the-art lattice calculation of the isovector quark transversity PDF $\delta u(x)-\delta d(x)$ of the proton, using the LaMET approach. 
The lattice matrix elements of the transversity quasi-PDF are calculated at four lattice spacings $a=\{0.098,0.085,0.064,0.049\}$~fm and various pion masses ranging between $220$ and $350$ MeV, with proton momenta up to $2.8$ GeV. The result is non-perturbatively renormalized in the hybrid scheme~\cite{Ji:2020brr} with self renormalization~\cite{LPC:2021xdx} proposed recently, which is a viable IR-safe renormalization approach, and extrapolated to the continuum, physical mass and infinite momentum. We also make a comparison between our results and recent global analyses on the isovector quark transversity PDF of the proton~\cite{Cammarota:2020qcw,Gamberg:2022kdb}. %

{\em Theoretical Framework:}
The leading-twist quark transversity PDF $\delta{q}(x)$ of the proton is defined as~\cite{Jaffe:1991kp}, 
\begin{align}\label{eq:LCtransPDF}
\notag \delta q(x,\mu)=\int \frac{d\xi^-}{4\pi} e^{-ix P^+ \xi^-} &\langle PS_\perp|\bar{\psi}(\xi^-)\gamma^{+}\gamma^{\perp}\gamma_5  \\
&\times\mathcal{W}[\xi^-,0] \psi(0)|PS_\perp \rangle,
\end{align}
where $|PS_\perp\rangle$ denotes a transversely polarized proton with momentum $P$ along the $z$-direction and polarization $S_\perp$ along the transverse direction, $x$ is the momentum fraction carried by the quark, $\mu$ is the renormalization scale in the $\overline{\rm MS}$ scheme, $\xi^\pm=(\xi^t\pm \xi^z)/\sqrt{2}$ are light-cone coordinates and $\mathcal{W}[0,\xi^-]= \mathcal{P}\,{\rm exp}\big[ig\int_{\xi_-}^0 du ~ {n}\cdot A(un)\big]$ is the gauge link along the light-cone direction ensuring gauge invariance of the non-local quark bilinear correlator.

According to LaMET, we can extract the transversity PDF from the following transversity quasi-PDF on the lattice
\begin{align} \label{eq:quasitransPDF}
\delta\tilde{q}(x,P_z,1/a)&=N\int \frac{d z}{4 \pi} e^{ix z P_z } \tilde{h}(z,P_z,1/a), \\
\tilde{h}(z,P_z,1/a)&=\langle P S_\perp|\bar{\psi}(z) \gamma^{t} \gamma^{\perp}\gamma_5 \mathcal{W}[z,0] \psi(0)|P S_\perp \rangle,  \nn
\end{align}
where $\tilde{h}(z,P_z,1/a)$ is the equal-time or quasi-light-front (quasi-LF) correlation that can be calculated directly on the lattice, $a$ denotes the lattice spacing, and $N=P_z/P_t$ is a normalization factor. Throughout the paper, we take the flavor combination $\delta\tilde{u}(x)-\delta\tilde{d}(x)$ to eliminate disconnected contributions. 

The bare quasi-LF correlation above contains both linear and logarithmic ultraviolet (UV) divergences, which need to be removed by a proper non-perturbative renormalization. 
Various approaches have been suggested and implemented in the literature~\cite{Chen:2016fxx,Izubuchi:2018srq,Alexandrou:2017huk,Radyushkin:2018cvn,Braun:2018brg,Li:2018tpe}. However, they all suffer from the problem that the renormalization factor introduces undesired non-perturbative contributions distorting the IR behavior of the original quasi-LF correlation. This is avoided in the so-called hybrid scheme~\cite{Ji:2020brr}, where one separates the quasi-LF correlations at short and long distances and renormalize them separately. At short distances, the renormalization is done by dividing by the same hadron matrix element in the rest frame, as is done in the ratio scheme~\cite{Radyushkin:2018cvn}; while at long distances one removes the UV divergences of the quasi-LF correlation only, which are determined by the so-called self renormalization~\cite{LPC:2021xdx} through fitting the bare matrix elements at multiple lattice spacings to a physics-dictated functional form. 
The renormalized quasi-LF correlation then takes the following form
\begin{align}\label{eq:hybridscheme}
\tilde{h}_R(z,P_z)
=&\frac{\tilde{h}(z,P_z,1/a)}{\tilde{h}(z,P_z=0,1/a)}\theta(z_s-|z|)\nn\\
&+	\eta_s\frac{\tilde{h}(z,P_z,1/a)}{Z_R(z,1/a)} \theta(|z|-z_s), 
\end{align}
where $z_s$ is introduced to separate the short and long distances, and has to lie in the perturbative region. $Z_R(z,1/a)$ denotes the renormalization factor extracted from the self renormaliztaion procedure using the same transversity three-point correlator in the rest frame. Details are given in the Supplemental Material~\cite{supp}. We have included a factor $\eta_s$ which is similar to a scheme conversion factor and determined by requiring continuity of the renormalized quasi-LF correlation at $z=z_s$. 
Of course, one needs to check the stability of the final result with respect to the choice of $z_s$. We vary $z_s$ in a certain perturbative range and include the difference in the systematic uncertainties. Note that all singular dependence on $a$ has been canceled in the above equation so that $\tilde h_R$ is independent of $a$ (up to some finite discretization effects).  

\begin{figure}[thbp]
\includegraphics[width=.4\textwidth]{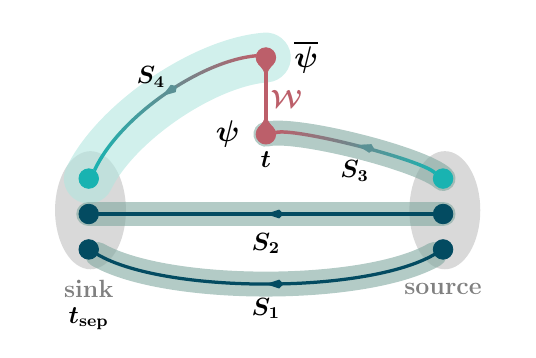}
\caption{Illustration of the sequential source method. The time direction is from source to sink. Propagators $S_{1,2}$ are combined to construct the sequential source. The inversion with the sequential source gives propagator $S_4$. $S_4$, $S_3$, gauge link $\mathcal{W}$ and necessary projectors are assembled to get the three-point correlator.} 
\label{fig:lattsetup}
\end{figure}

After performing a Fourier transform to momentum space, we can match the transversity quasi-PDF to the transversity PDF through the following perturbative matching 
\begin{align} \label{eq:matching}
\delta\tilde{q}(x, P_z) =& \int_{-1}^1{dy \over |y|} C\left({x\over y},{\mu\over yP_z}\right) \delta q(y,\mu)\nn\\
&+\mathcal{O}\Big({\Lambda^2_{\rm{QCD}}\over({{x}}P_z)^2},{\Lambda^2_{\rm{QCD}}\over((1-{{x}})P_z)^2}\Big)\,,
\end{align}
where $C$ is the perturbative matching kernel whose explicit expression to $O(\alpha_s)$ is given in the Supplemental Material~\cite{supp}. The transversity at negative $y$ can be interpreted as the antiquark transversity via the relation $\delta \bar{q}(y,\mu)=-\delta q(-y,\mu)$.  $\mathcal{O}\Big({\Lambda^2_{\rm{QCD}}\over(xP_z)^2},{\Lambda^2_{\rm{QCD}}\over((1-x)P_z)^2}\Big)$ denotes higher-twist contributions suppressed by the nucleon momentum $P_z$.


{\em Lattice calculation:}
To improve the signal-to-noise ratio of calculations with high-momentum nucleon states, we employ the momentum smearing source technique~\cite{Bali:2016lva}. {Besides, we apply two steps APE smearing to further improve the signal.} We also use the sequential source method with fixed sink to calculate the quark three-point correlator, as illustrated in Fig.~\ref{fig:lattsetup}.  
In order to perform the continuum and chiral extrapolation, we use six different lattice ensembles generated by the CLS collaboration~\cite{Bruno:2014jqa} with lattice spacing $a=\{0.098,0.085,0.064,0.049\}~\mathrm{fm}$ and pion masses ranging between $354$ and $222$ MeV. For each ensemble, we calculate several source-sink separations with hundreds to thousands of measurements among 500 gauge configurations (for X650 it is {1500} because the autocorrelation time is larger for this rather coarse lattice). Details of the lattice setup and parameters are collected in Table~\ref{Tab:setup}. 

The bare matrix elements are calculated with the nucleon carrying different spatial momenta on each ensemble: $P_z=\{1.84,2.37,2.63\}$ GeV on X650, $P_z=\{1.82,2.27,2.73\}$ GeV on H102, $P_z=1.82$ GeV on H105 and C101, $P_z=\{1.62,2.02,2.43,2.83,3.24\}$ GeV on N203 and $P_z=\{2.09,2.62\}$ GeV on N302. We also calculate the zero-momentum bare matrix elements for each ensemble, which provide the renormalization factor at short distance and the inputs to extract the renormalization factor at long distance through self renormalization. 

\begin{table}
\centering
\begin{tabular}{cclccccccc}
\hline
\hline
Ensemble ~&$a$(fm) ~& \ \!$L^3\times T$  ~& $m_\pi$(MeV) ~& $m_\pi L$  & $N_\mathrm{conf.}$\\
\hline
X650  ~& 0.098  ~& $48^3\times 48$  ~& 338   ~&8.1    &{{{1500}}}                                \\ 
H102  ~& 0.085  ~& $32^3\times 96$  ~& 354   ~&4.9    &500                 \\ 
H105  ~&        ~& $32^3\times 96$   ~& 281  ~&3.9   &500              \\ 
C101  ~&        ~& $48^3\times 96$   ~& 222  ~&4.6   &500              \\ 
N203  ~& 0.064  ~& $48^3\times 128$  ~& 348  ~&5.4   &500      \\ 
N302  ~& 0.049  ~& $48^3\times 128$  ~& 348  ~&4.2   &500        \\
\hline
\end{tabular}
 \caption{The simulation setup, including lattice spacing $a$, lattice size $L^3\times T$ and the pion masses~\cite{RQCD:toappear}.  {For more details we refer the interested readers to Table.~I in the Supplemental Material~\cite{supp}.}
 For zero momentum matrix elements, the numbers of configurations used are {reduced to 350 for H102, and 100 for H105 and N203}, because that already gives a satisfactory signal-to-noise ratio.}
 \label{Tab:setup}
\end{table}

To extract the ground state matrix element, we decompose the two-point correlator $C^{2\text{pt}}(P_z,t_\text{sep})$ and three-point correlator $C_\Gamma^{3\text{pt}} (P_z, t, t_\text{sep})$ (with $\Gamma=\gamma^t\gamma^\perp \gamma_5$) as in~\cite{Bhattacharya:2013ehc}, 
%
\begin{align}\label{eq:twostate}
C^\text{2pt}(P_z,t_\text{sep}) &=|{\cal A}_0|^2 e^{-E_0t_\text{sep}}+|{\cal A}_1|^2 e^{-E_1t_\text{sep}}+\cdots\,,\nn\\
C^\text{3pt}_{\Gamma}(P_z,t,t_\text{sep}) &=
   |{\cal A}_0|^2 \langle 0 | {O}_\Gamma | 0 \rangle  e^{-E_0t_\text{sep}} \nonumber\\
   &+|{\cal A}_1|^2 \langle 1 | {O}_\Gamma | 1 \rangle  e^{-E_1t_\text{sep}} \nonumber\\
   &+{\cal A}_1{\cal A}_0^* \langle 1 | {O}_\Gamma | 0 \rangle  e^{-E_1 (t_\text{sep}-t)} e^{-E_0 t} \nonumber\\
   &+{\cal A}_0{\cal A}_1^* \langle 0 | {O}_\Gamma | 1 \rangle  e^{-E_0 (t_\text{sep}-t)} e^{-E_1 t} + \cdots \,,
\end{align}
where $\left\langle 0 \left| \mathcal{O}_\Gamma \right|  0\right\rangle=\tilde{h}(z,P_z,1/a)$ is the ground state matrix element, $t$ denotes the insertion time of $O_\Gamma$. The ellipses denote the contribution from higher excited states of the nucleon which decay faster than the ground state and first-excited state. We extract the ground state matrix element by performing a two-state combined fit with $C^{2\text{pt}}(P_z,t_\text{sep})$ and the ratio  $R_\Gamma(z,P_z,t_\text{sep},t)={C_\Gamma^{3\text{pt}}(z,P_z,t_\text{sep},t)}/{C^{2\text{pt}}(P_z,t_\text{sep})}$. 
Details of the fit can be found in the Supplemental Material~\cite{supp}.

\begin{figure*}[thbp]
\includegraphics[width=1\textwidth]{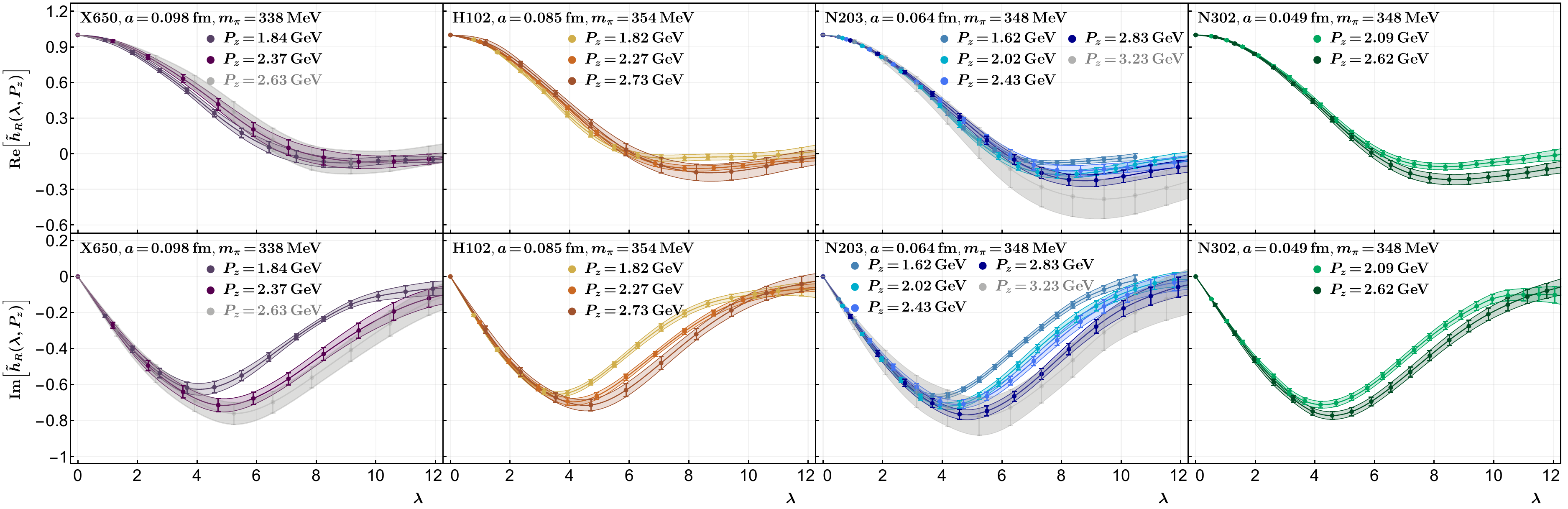}
\vspace{-1em}
\caption{The real (top) and imaginary (bottom) parts of the renormalized matrix elements for different ensembles as functions of $\lambda$ at scale {$\mu=2$~GeV}.} 
\label{fig:RenorME-tsep}
\end{figure*}

\begin{figure}[htbp]
\includegraphics[width=.46\textwidth]{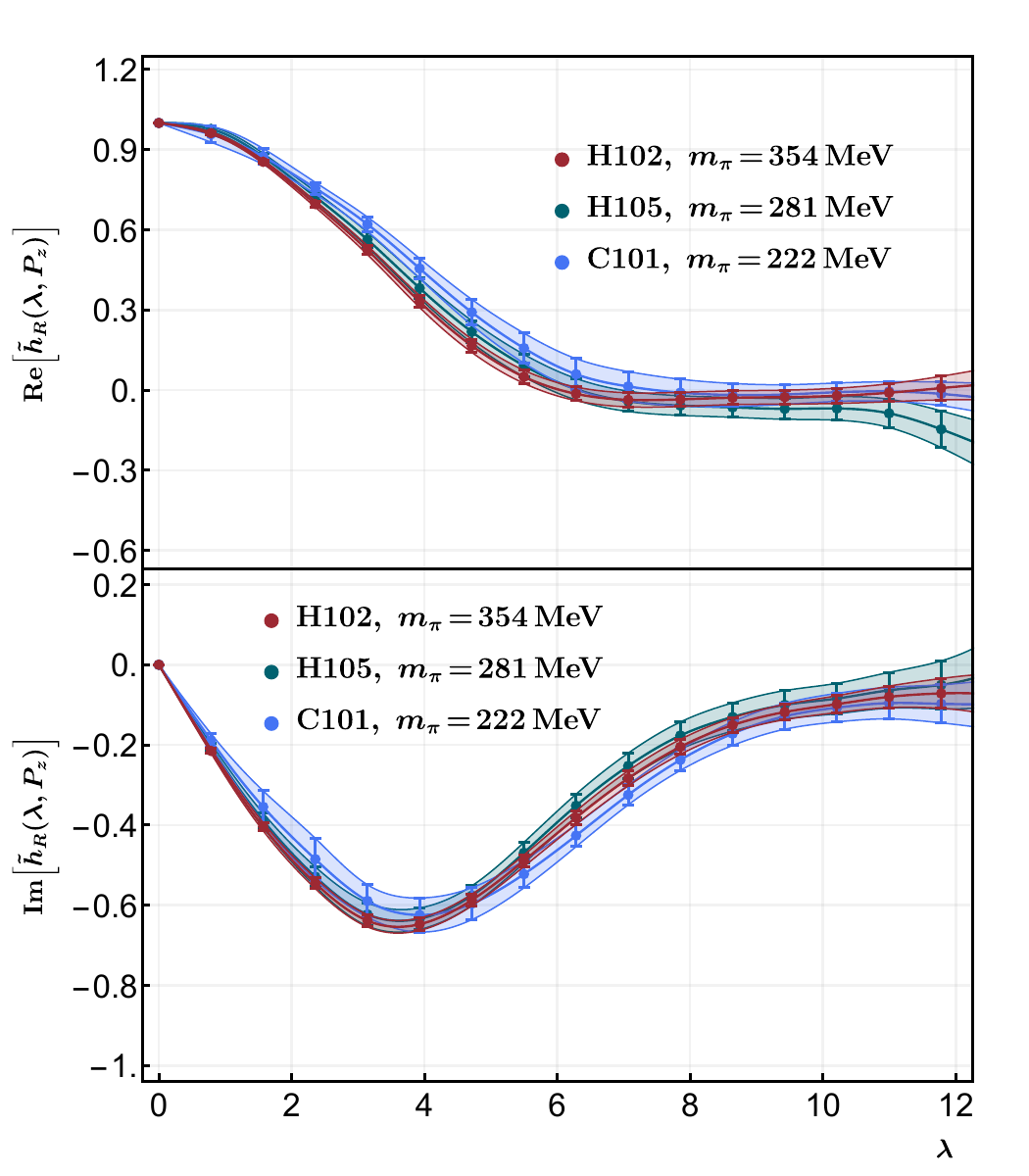}
\caption{The pion mass dependence of the real (top) and imaginary (bottom) part of the renormalized matrix elements for different ensembles with the same lattice spacing $a=0.085$~fm and momentum $P_z=1.82$~GeV.} 
\label{fig:RenorME-pionmass}                                    
\end{figure}

In Fig.~\ref{fig:RenorME-tsep}, we plot the renormalized quasi-LF correlation in the hybrid scheme, as a function of the quasi-LF distance $\lambda=zP_z$. 
The data points are obtained on ensembles with nearly the same pion mass. As can be seen from the figure, the results show a good convergence as $P_z$ increases for all ensembles. 
However, as we increase the nucleon boost momentum, the excited-state contamination worsens so that uncertainties increase. 
The $P_z=2.63$~GeV data on X650 and 
$P_z=3.23$~GeV data on N203 have much larger uncertainties compared to other data sets due to the relatively larger momentum on these ensembles.  
We exclude them in our analysis below. In Fig.~\ref{fig:RenorME-pionmass}, we show the pion mass dependence of the renormalized quasi-LF correlations on ensembles with the same lattice spacing $a=0.085$~fm, where the pion masses are $m_\pi=\{354,281,222\}$ MeV, respectively. As shown in the figure, the results only exhibit a very mild dependence on the pion mass. 




From the figures above, we can see that the uncertainty of the renormalized quasi-LF correlation grows at large distance, while a Fourier transform to momentum space requires the quasi-LF correlation at all distances. If we do a brute-force truncation and Fourier transform,  unphysical oscillations will appear in the momentum space distribution. To resolve this issue, we adopt a physics-based extrapolation form~\cite{Ji:2020brr} at large quasi-LF distance
\begin{align}
	  H^{\rm R}_{m}(z, P_z) &= \Big[\frac{c_1}{(i\lambda)^a} + e^{-i\lambda}\frac{c_2}{(-i \lambda)^b}\Big]e^{-\lambda/\lambda_0},
	  \label{eq:extrap}
\end{align} 
where the algebraic terms in the square bracket account for a power law behavior of the transversity PDFs in the endpoint region, and the exponential term comes from the expectation that at finite momentum the correlation function has a finite correlation length (denoted as $\lambda_0$)~\cite{Ji:2020brr}, which becomes infinite when the momentum goes to infinity. The detail of the extrapolation is expected to affect the final results in the small and large $x$ region where LaMET expansion breaks down~\cite{Ji:2020ect}. An example of the extrapolation is given in the Supplemental Material~\cite{supp}.

%
After the renormalization and extrapolation, we can Fourier transform the quasi-LF correlation to momentum space and extract the transversity PDF by applying perturbative matching. 
The transversity PDF extracted this way still contains lattice artifacts which shall be removed by performing a continuum extrapolation. Also, our calculations are not done at infinite momentum and the physical point, so we have to extrapolate to infinite momentum and physical pion mass. 
To this end, we perform a simultaneous extrapolation using the following functional form including $a$, $P_z$, and  $m_{\pi}$, 
\begin{align}
\delta q\left(x,P_z,a, m_\pi\right)&=\frac{1-g' m_\pi^2\,\ln (m_\pi^2/\mu_0^2)+m_\pi^2 k(x)}{1-g' m_\pi^2\,\ln (m_\pi^2/\mu_0^2)}\nn\\
&\hspace{-4em}\times\left[\delta q_0(x)+a^2f(x)+a^2P_z^2 h(x)+\frac{g(x,a)}{P_z^2}\right],\label{eq:combinedfit}
\end{align}
where $\delta q(x,P_z,a, m_\pi)$ on the l.h.s. denotes the transversity PDF results obtained in our calculation on different ensembles. The extrapolation of the pion mass dependence follows from the study in Ref.~\cite{Chen:2001eg} (with $\mu_0=1$ GeV and $g'=-(4g_A^2+1)/[2(4\pi f_\pi)^2]$ and $g_A$ the axial charge of the nucleon) and for the CLS ensembles in Ref.~\cite{RQCD:2019hps}.
The $a^2 f(x)$ term denotes the leading discretization error which begins at $\mathcal{O}(a^2)$ as the lattice action is already $\mathcal{O}(a)$ improved.  The $a^2 P_z^2$ term represents the momentum-dependent discretizaton error. The {last} term in the square bracket accounts for the leading higher-twist contribution, where in the numerator we also include a potential $a$-dependence. Choosing $g(x,a)$ as a generic function or parametrizing it with a quadratic form in $a$ 
yields similar results with slightly different errors. We refer the readers to the Supplemental Material~\cite{supp} for details of this extrapolation. Eventually the desired transversity PDF is given by
\begin{align}
\delta q\left(x\right)&\!=\!\frac{1\!-\!g' m_{\pi,\rm phys}^2\ln (m_{\pi,\rm phys}^2/\mu_0^2)\!+\!m_{\pi,\rm phys}^2 k(x)}{1\!-\!g' m_{\pi,\rm phys}^2\,\ln (m_{\pi,\rm phys}^2/\mu_0^2)}\delta q_0(x),
\end{align}
where $m_{\pi,\rm phys}=135$ MeV.

\begin{figure}[thbp]
\includegraphics[width=.46\textwidth]{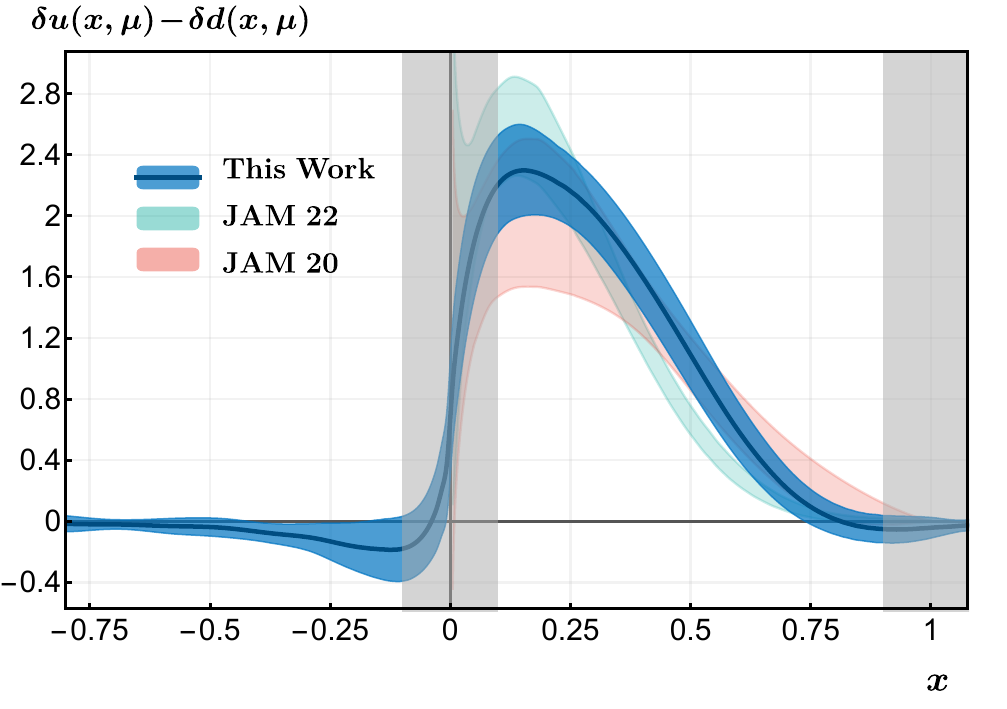}
\caption{Our final proton isovector transversity PDF at renormalization scale $\mu=2$~GeV, extrapolated to the continuum, physical pion mass and infinite momentum limit ($a\rightarrow0$, $m_\pi\rightarrow m_{\pi,{\rm phys}}$, $P_z\rightarrow\infty$), compared with JAM20~\cite{Cammarota:2020qcw} and JAM22~\cite{Gamberg:2022kdb} global fits. {All results are normalized to the nucleon isovector tensor charge $g_T$. The blue error band includes both statistical and systematic errors. {The vertical light gray bands denote the endpoint regions where LaMET predictions are not reliable.}}}\label{fig:mtchdPDF}
\end{figure}

{\em Numerical result:}
Our final result for the physical isovector quark transversity PDF $\delta u(x)-\delta d(x)$ {(normalized by nucleon isovector tensor charge $g_T$)} 
is shown in Fig.~\ref{fig:mtchdPDF}, together with the results of recent global analyses from the JAM collaboration (JAM20~\cite{Cammarota:2020qcw} and JAM22~\cite{Gamberg:2022kdb}). JAM20 is the global analysis that finds, for the first time, agreement between phenomenology and lattice calculation of all nucleon tensor charges $\delta u, \delta d$ and the isovector combination $g_T$.  
JAM22 provides an update to JAM20 by including certain new data sets and constraints from the Soffer bound and lattice $g_T$ result. Given the sensitivity of the results to the choice of data sets, we tend to view the difference between the two curves as an indication of systematic uncertainties from global fits. As can be seen from the figure, our result lies between the two global analyses and agrees with both within $1\sim2\sigma$ error. Note that we have plotted two shaded bands at the endpoint regions to indicate that LaMET predictions are not reliable there (taken as $x\in [-0.1,0.1]$ and $x\in [0.9, 1]$), which are estimated from the observation that the higher-twist terms in LaMET factorization Eq.~\eqref{eq:matching} are of $\mathcal{O}(1)$, {\it{i.e.,}} ${\Lambda_{\mathrm{QCD}}/(xP_z)}\sim 1$ and ${\Lambda_{\mathrm{QCD}}/((1-x)P_z)}\sim 1$ with the highest momentum in this work $P_z=2.83\mathrm{GeV}$. This estimation is consistent with the recent approach incorporating the renormalization group resummation (RGR) procedure~\cite{Su:2022fiu}, as can be seen from the Supplemental Material~\cite{supp}. Our error band includes both statistical and systematic uncertainties, where the latter have four different sources. 
The first is the renormalization scale dependence, which is estimated by varying the scale to 3 GeV. 
The second error comes from the extrapolation to continuum, infinite momentum, and physical mass, which is relatively small. The third error is from the choice of $z_s$ in the hybrid renormalization scheme. We choose $z_s=0.3$~fm, vary it down to 
$z_s=0.18$~fm and take the difference as a systematic error. Lastly, the large $\lambda$ extrapolation also introduces some error that mainly affects the small-$x$ region $-0.2\lesssim x\lesssim0.2$. We have chosen different regions to do the extrapolation to estimate this error. 
More details of the systematic uncertainties can be found in the Supplemental Material~\cite{supp}.

In the negative $x$ region, our result is consistent with zero, which puts a strong constraint on the sea flavor asymmetry. 
JAM plans to update their global analysis by including spin asymmetry data from STAR~\cite{Gamberg:2022kdb}. It will be very interesting to see how their new analysis compares with our lattice result.

{\em Summary:}
We present a state-of-the-art calculation of the isovector quark transversity PDF with LaMET. The calculation is done at various lattice spacings, pion masses and large nucleon momenta, and extrapolated to the continuum, physical mass and infinite momentum limit. With high statistics, we have performed multi-state analyses with multiple source-sink separations to remove the excited-state contamination, and applied state-of-the-art renormalization, matching and extrapolation. Our result provides the most reliable lattice prediction of the isovector quark transversity PDF in the proton so far, and will offer guidance to measurements at JLab and EIC.

\begin{acknowledgments}
\section*{Acknowledgments}
We thank the CLS Collaboration for sharing the lattices used to perform this study. We are grateful to Wolfgang S\"oldner for helpful discussions on X650 ensemble, and Daniel Pitonyak and Nobuo Sato for providing the JAM fits data of the transversity PDF. We also thank Yu-Sheng Liu and Yong Zhao for helpful correspondence. AS thanks the University of the Basque Country, Bilbao for hospitality. This work was supported in part by the High Performance Computing Center of Central South University. The authors gratefully acknowledge the Gauss Centre for Supercomputing e.V. (www.gauss-centre.eu) for funding this project by providing computing time on the GCS Supercomputer SuperMUC at Leibniz Supercomputing Centre (www.lrz.de). The LQCD calculations were performed using the multigrid algorithm~\cite{Babich:2010qb,Osborn:2010mb} and Chroma software
suite~\cite{Edwards:2004sx}. XNX is supported in part by the National Natural Science Foundation of China
under Grant No.~11905296.
YY is supported in part by the Strategic Priority Research Program of Chinese Academy of Sciences, Grant No. XDB34030303 and XDPB15. FY, LM and JHZ are supported in part by the National Natural Science Foundation of China under Grant No. 11975051. AS, HTS, PS, YY and JHZ are also supported by a NSFC-DFG joint grant under grant No. 12061131006 and SCHA~458/22. PS is also supported by Strategic Priority Research Program of the Chinese Academy of Sciences under grant number XDB34030301. X.J. is supported by the U.S. Department of Energy, Office of Science, Office of Nuclear Physics, under contract number DE-SC0020682. JWC is supported by the Taiwan Ministry of Science and Technology under Grant No. 111-2112-M-002-017-  and the Kenda Foundation.
\end{acknowledgments}

\bibliographystyle{apsrev}
\bibliography{transversity}

\clearpage

\section*{Supplemental Material}\label{sec:supp}

\subsection{Additional information on renormalization and matching}

\subsubsection{Self renormalization on the CLS ensembles}
\begin{figure}[thbp]
\includegraphics[width=.45\textwidth]{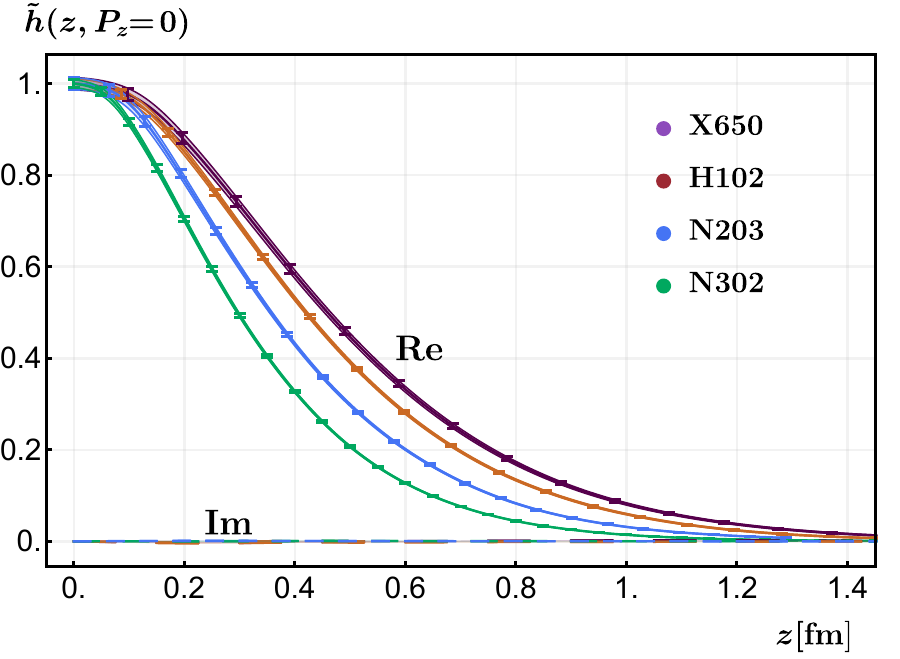}
\caption{Bare matrix elements in the rest frame. Their imaginary parts are consistent with zero. }\label{fig:BaMtrxElmntPz0Fig}
\end{figure}

In this section, we give more details on the hybrid renormalization procedure used in this work. In the hybrid scheme, the short- and long-distance quasi-LF correlations $\tilde{h}(z,P_z,1/a)$ are renormalized separately. At short distances $|z|<z_s$, we take the inverse of the nucleon matrix element in the rest frame $[\tilde{h}(z,P_z=0,1/a)]^{-1}$ as the renormalization factor. The bare matrix element in the rest frame is shown in Fig.~\ref{fig:BaMtrxElmntPz0Fig}, where the imaginary part is consistent with $0$. The renormalized matrix element takes the following form
\begin{align}\label{eq:HYBshort}
    \tilde{h}_R(z,P_z) = \frac{\tilde{h}(z,P_z,1/a)}{\tilde{h}(z,P_z\!=\!0,1/a)},\;\; |z|<z_s.
\end{align}

\begin{figure}[htbp]
\includegraphics[width=.45\textwidth]{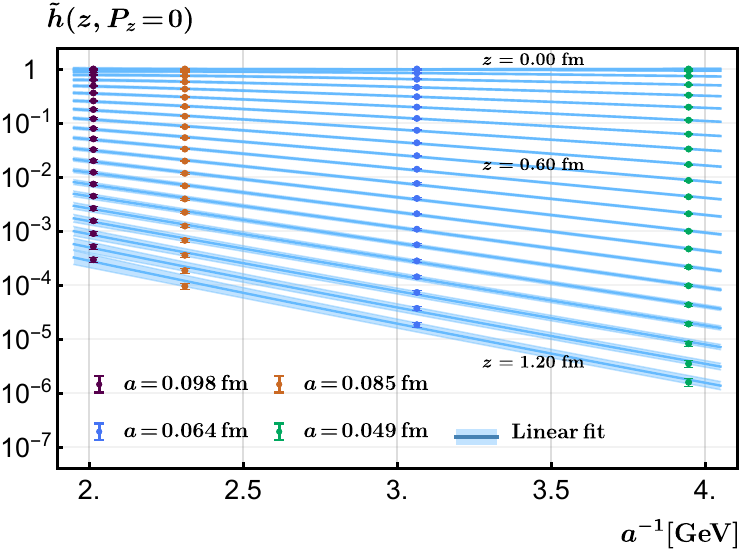}
\caption{Fit of the bare nucleon transversity matrix elements in the rest frame. Colorful points represent the bare matrix elements from lattice calculation and blue bands are fitted values using Eq.~(\ref{eq:lnbrMEFit}). The parameters $k$ and $\Lambda_{\mathrm{QCD}}$ are fitted to be $k=4.356~\mathrm{GeV}^{-1}\mathrm{fm}^{-1}$ and $\Lambda_{\mathrm{QCD}}=0.1~\mathrm{GeV}$. }
\label{fig:lnBrMtrxElmntPz0}
\end{figure}
\begin{figure}[htbp]
\includegraphics[width=.45\textwidth]{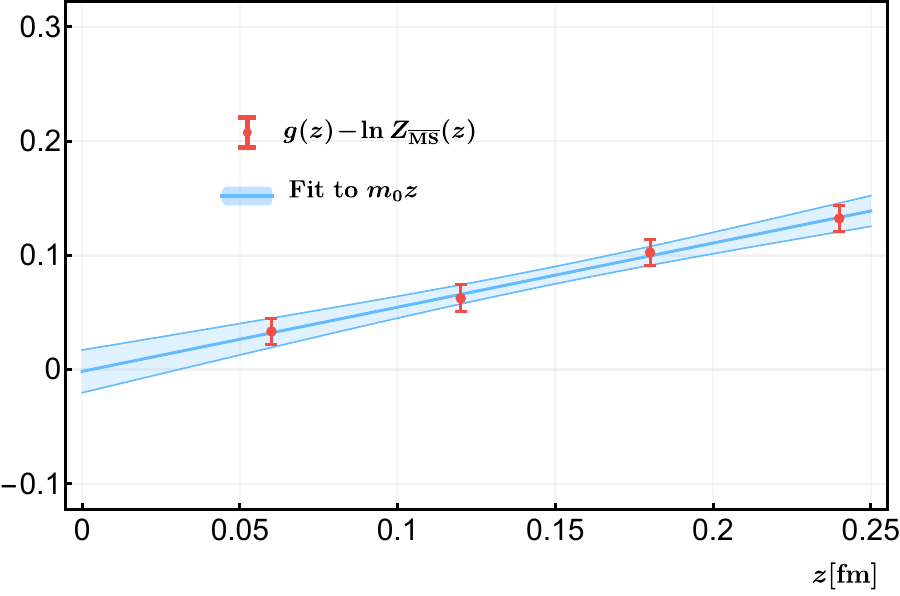}
\caption{The fit of $m_0$. Red points are $g(z)-\ln Z_{\overline{\rm{MS}}}(z)$ in the small-$z$ region. The blue band is the fit to $m_0 z + b$, where we tune the parameter $d$ in Eq.~(\ref{eq:lnbrMEFit}) to minimize $|b|$. The fit gives $m_0=0.56\,\mathrm{fm}^{-1}$, $d=-0.663$ and $b=-0.0015$.}
\label{fig:fitm0}
\end{figure}
\begin{figure}[htbp]
\includegraphics[width=.48\textwidth]{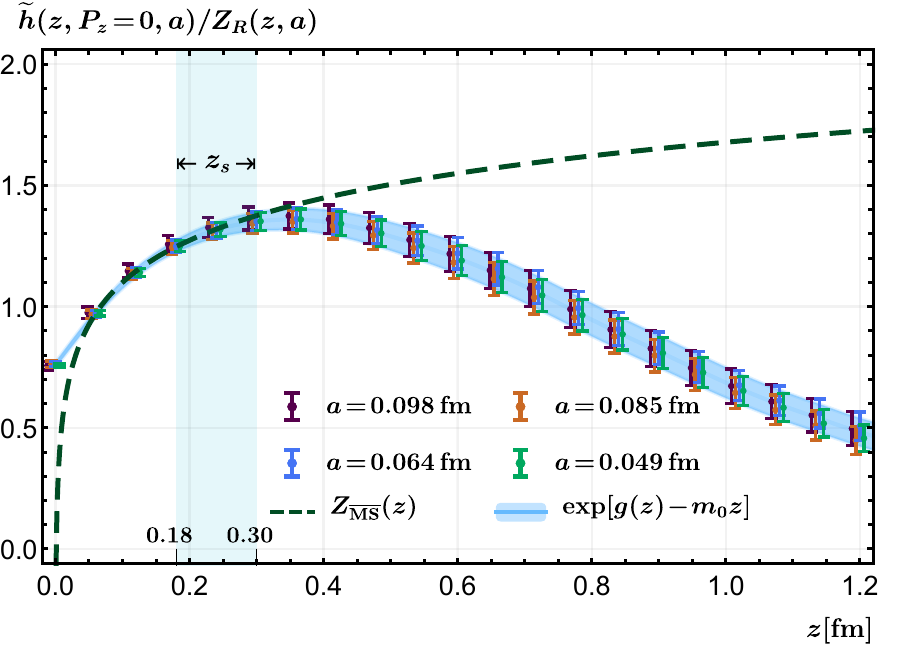}
\caption{The renormalized matrix element $\tilde h_R(z)$, for each $a$ (colorful points) and the fitted result (blue band). We have slightly shifted the data points to the $\pm z$ direction for clarity. The renormalized matrix elements overlap nicely with the perturbative one-loop result $Z_{\overline{\mathrm{MS}}}(z)$ at short distances, except at very small $z$ where discretization effects and higher-order perturbative corrections become important. }\label{fig:ZR}
\end{figure}

At long distances $|z|>z_s$, 
the renormalized matrix element is given by
\begin{align}\label{eq:HYBlong}
    \tilde{h}_R(z,P_z,1/a) =\eta_s \frac{\tilde{h}(z,P_z,1/a)}{Z_R(z,1/a)},\;\; |z|>z_s,
\end{align}
where we have included a factor $\eta_s=Z_R(z_s,1/a)/\tilde{h}(z_s,P_z\!=\!0,1/a)$ which is similar to a scheme conversion factor and guarantees continuity of the renormalized matrix element at $z=z_s$.
The self renormalization factor $Z_R(z,1/a)$ is obtained by 
fitting the bare matrix elements in the rest frame to the following perturbative-QCD-dictated functional form~\cite{LPC:2021xdx}
\begin{align}\label{eq:lnbrMEFit}
    &\ln \tilde{h}(z,1/a) = \frac{kz}{a\ln(a\Lambda_{\rm{QCD}})}\!+\!g(z)\!+\!f(z)a^2\!\notag\\
    &+\!\frac{3C_F}{11-2N_f/3}\ln\left[\frac{\ln{[1/(a\Lambda_{\rm{QCD}})]}}{\ln{[\mu/\Lambda_{\rm{QCD}}]}}\right]\!+\!\ln\left[1\!+\!\frac{d}{\ln(a\Lambda_{\rm{QCD}})}\right],\end{align}
where the first term on the r.h.s. is the linear divergence, $g(z)=g_0(z)+m_0 z$ contains the intrinsic non-perturbative physics $g_0(z)$ we are interested in and a renormalon ambiguity term $m_{0} z$, $f(z) a^2$ accounts for the discretization effects. The last two terms come from the resummation of leading and sub-leading logarithmic divergences, which only affect the overall normalization at different lattice spacings. {To partially account for higher-order perturbative effects as well as remaining lattice artifacts, we also treat $d$ and $\Lambda_{\rm QCD}$ as fitting parameters}~\cite{LPC:2021xdx}. The renormalization factor and the renormalized matrix element are then given by
\begin{align}\label{eq:physcalfit}
Z_R(z,1/a) &= \frac{\tilde{h}_(z,1/a)}{\tilde{h}_{\mathrm{R}}(z)},\nn\\
\tilde{h}_{\mathrm{R}}(z)&= \exp[g(z)-m_0 z]=\exp [g_0(z)],
\end{align}
with $\tilde{h}_{\mathrm{R}}(z)$ being required to be equal to the continuum perturbative $\overline{\rm MS}$ result at short distances, which reads at one-loop 
\begin{align}\label{eq:oneloopPZ0}
Z_{\overline{\mathrm{MS}}}(z)=1+
\frac{\alpha_s C_F}{2\pi}\left(2\ln\left(z^2\mu^2e^{2\gamma_E}/4\right)+2\right). 
\end{align}
The fitting of the bare matrix element and $m_0$ is shown in Figs.~\ref{fig:lnBrMtrxElmntPz0} and \ref{fig:fitm0}, respectively. The comparison of the renormalized matrix element with the perturbative one-loop $\overline{\rm MS}$ result is given in Fig.~\ref{fig:ZR}. As can be seen from the latter figure, the agreement between the renormalized matrix element and the continuum one-loop $\overline{\rm MS}$ result is good at short distances, except at very small $z$ where higher-order corrections get important. In our analysis, we choose $z_s=0.3~\mathrm{fm}$ and vary it down to $0.18$~fm (plotted as shaded region in Fig.~\ref{fig:ZR}) to account for systematic uncertainties related to the choice of $z_s$.

\begin{widetext}

\subsubsection{One-loop matching in coordinate and momentum space}
In this section, we present the relevant one-loop matching kernel both in coordinate and in momentum space. Our calculation is done in Feynman gauge. Similar calculations have also been done in Refs.~\cite{Braun:2021gvv,Chou:2022drv}, which agree with our results.  
	
{\em Coordinate space:} We consider the transversity quasi-LF correlation with on-shell and massless external quark state
\begin{align} \label{eq:transPDFquasiLF}
\tilde{h}(z,p_z,\mu)&=\langle p|\bar{\psi}(z)  \gamma^{t} \gamma^{\perp}\gamma^5 \mathcal{W}[z,0] \psi(0)|p \rangle,
\end{align}
where the quark momentum is $p_\mu=(p_0,0,0,p_z)$, and dimensional regularization ($d=4-2\epsilon$) is used to regularize both UV and IR divergences. 
In coordinate space, factorization takes the following form
\begin{align}{}\label{eq:factorizationcoo}
\tilde{h}(z,\lambda=z p_z,\mu)=\int_{0}^{1} d\alpha  Z(\alpha,z^2\mu^2) h(\alpha \lambda,\mu) + h.t.,
\end{align}
where h.t. denotes higher-twist terms. At one-loop level, the matching kernel $Z(\alpha,z^2\mu^2)$ reads in the $\overline{\rm MS}$ scheme 
\begin{align}{}\label{eq:coomatchingkernelMSbar}
Z\left(\alpha,z^2\mu^2\right)=&\delta(1-\alpha)+\frac{\alpha_s C_F}{2\pi}\left\{-\left(\frac{2\alpha}{1-\alpha}\right)_{+} \left(\ln\frac{z^2\mu^2e^{2\gamma_E}}{4}+1 \right)- \left(\frac{4 \ln(1-\alpha)}{1-\alpha}\right)_{+} \right\} \theta(\alpha)\theta(1-\alpha) \notag\\
& + \frac{\alpha_s C_F}{2\pi} \left(2 \ln\frac{z^2\mu^2e^{2\gamma_E}}{4}+2\right)\delta(1-\alpha).
\end{align}
To obtain one-loop matching in the hybrid scheme, we begin with the ratio scheme. The quasi-LF correlation at zero momentum and short distance is given by
\begin{align}{}\label{eq:zeromom}
Z_0\left(z,\mu\right)=1+
\frac{\alpha_s C_F}{2\pi}\left(2\ln\frac{z^2\mu^2e^{2\gamma_E}}{4}+2\right).
\end{align} 
Thus, the one-loop matching in the ratio scheme is \begin{align}\label{eq:coomatchingkernelratio}
Z_r(\alpha,z^2\mu^2)=\delta\left(1-\alpha\right)+\frac{\alpha_s C_F}{2\pi}\left\{-\left(\frac{2\alpha}{1-\alpha}\right)_{+} \left(\ln\frac{z^2\mu^2e^{2\gamma_E}}{4}+1 \right)- \left(\frac{4 \ln(1-\alpha)}{1-\alpha}\right)_{+} \right\} \theta(\alpha)\theta(1-\alpha).
\end{align}
From this, we can easily obtain the one-loop matching kernel in the hybrid scheme as
\begin{align}\label{eq:coomatchingkernelhybrid}
Z_h\left(\alpha,z^2\mu^2,\frac{z^2}{z_s^2}\right)
&=Z_r\left(\alpha,z^2\mu^2\right)+\frac{\alpha_s C_F}{\pi}  \ln\left(\frac{z^2}{z_s^2}\right) \delta\left(1-\alpha\right) \theta\left(|z|-z_s\right),
\end{align}	
which reduces to the ratio scheme matching when $z_s \rightarrow \infty$.

{\em Momentum space:} The transversity quasi-PDF is defined as a Fourier transform of the quasi-LF correlation
\begin{equation}\label{eq:FTquasi}
\delta\tilde{q}\left(x,\frac{\mu}{p_z}\right)= \int_{-\infty}^{\infty} \frac{d\lambda}{2\pi}e^{ix\lambda}\tilde{h}\left(\lambda,\frac{\mu^2\lambda^2}{p_z^2}\right).
\end{equation}  
Thus, the matching kernel in momentum space is related to that in coordinate space by a double Fourier transform 
\begin{equation}\label{eq:FTmatchingkernel}
C\left(x,\frac{\mu}{p_z}\right)= \int_{-\infty}^{\infty} \frac{d\lambda}{2\pi}e^{ix\lambda}\int_{-1}^{1}d\alpha e^{-i\alpha\lambda} Z\left(\alpha,\frac{\mu^2\lambda^2}{p_z^2}\right).
\end{equation}  
The factorization formula in momentum space has been given in Eq.~(4\vphantom{\ref{eq:matching}}) of the main text.  Starting with Eq.~(\ref{eq:coomatchingkernelratio}) and using Eq.~(\ref{eq:FTmatchingkernel}) together with the factorization formula, we have the following result in the ratio scheme
\begin{equation}\label{eq:mommatchingkernelratio}
 C_r\left(x,\frac{\mu}{p_z}\right)=\delta\left(1-x\right)+\frac{\alpha_s C_F}{2\pi}\begin{cases}
\left[\frac{2x}{1-x}\ln\frac{x}{x-1}-\frac{2}{1-x}\right]_+ & x > 1 \\
\left[\frac{2x}{1-x}\left(\ln\frac{4p_z^2}{\mu^2}+\ln x (1-x)\right)+2\right]_+ & 0<x<1 \\
\left[-\frac{2x}{1-x}\ln\frac{x}{x-1}+\frac{2}{1-x}\right]_+ & x<0 ,
\end{cases}
\end{equation}
whereas in the hybrid scheme it becomes 
\begin{align}\label{eq:mommatchingkernelhybrid}
C_h\left(x,\frac{\mu}{p_z},\lambda_s\right)&= C_r\left(x,\frac{\mu}{p_z}\right)+\delta C\left(x,\frac{\mu}{p_z},\lambda_s\right)
=C_r\left(x,\frac{\mu}{p_z}\right)+\frac{\alpha_s C_F}{\pi} \left[-\frac{1}{|1-x|}+\frac{2 \mathrm{Si}((1-x)\lambda_s)}{\pi(1-x)}\right]_+ .
\end{align}


\end{widetext}

\subsection{Lattice data analysis}

\subsubsection{Extraction of the ground state information}

\begin{figure}[htbp]
\includegraphics[width=0.46\textwidth]{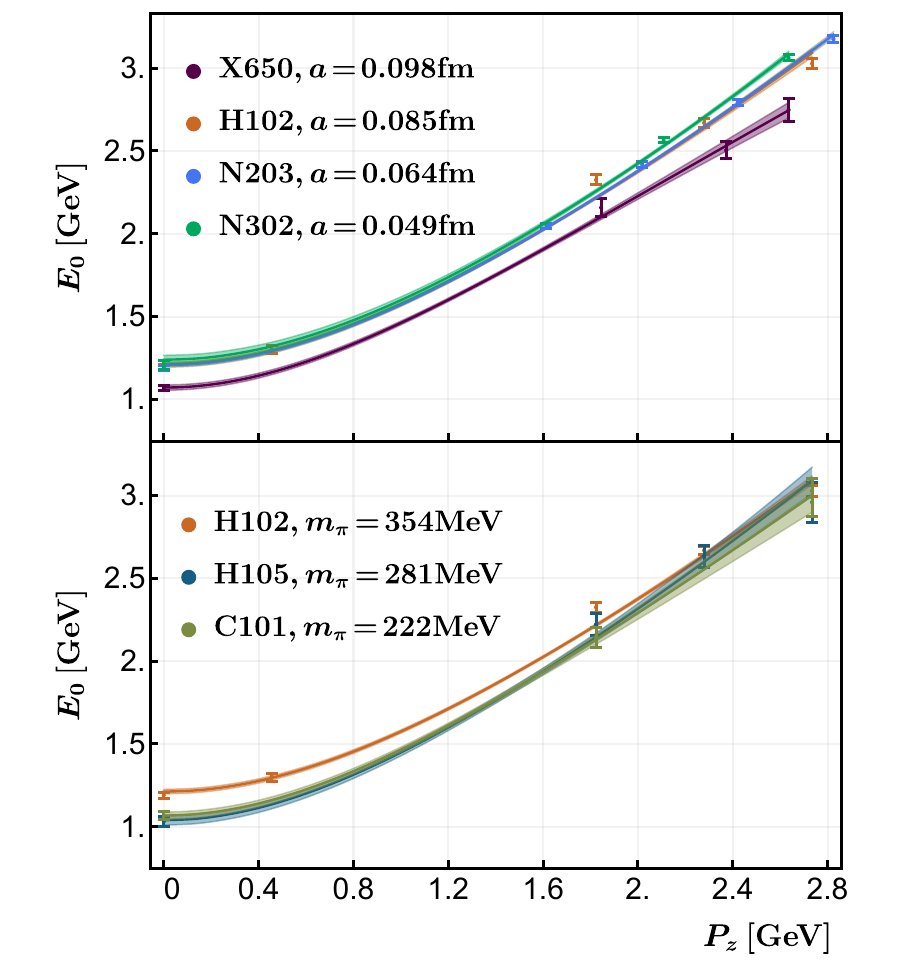}
\caption{The dispersion relation of CLS ensembles at four different lattice spacings used in this work. In the top subfigure we compare ensembles with different lattice spacing but roughly the same $\pi$ mass: $m_\pi\approx 340\mathrm{MeV}$, while in the bottom subfigure we compare ensembles with the same lattice spacing $a=0.085\,\mathrm{fm}$ but different $m_{\pi}$.}\label{fig:dsprtn}
\end{figure}

We show the dispersion relation of each ensemble in Fig.~\ref{fig:dsprtn}. The effective mass can be extracted by fitting the two-point correlation function. We fit the effective mass using 
\begin{equation}
 E(P_z)=\sqrt{m^2+P_z^2+c^2\, a^2\, P_z^4},   
\end{equation}
where a quadratic term in lattice spacing $a$ is included to parameterize the discretization error. We find that the extracted effective masses are consistent with the dispersion relation within $3\sigma$ error.

\begin{figure*}[htbp]
\begin{center}
\includegraphics[width=1\textwidth]{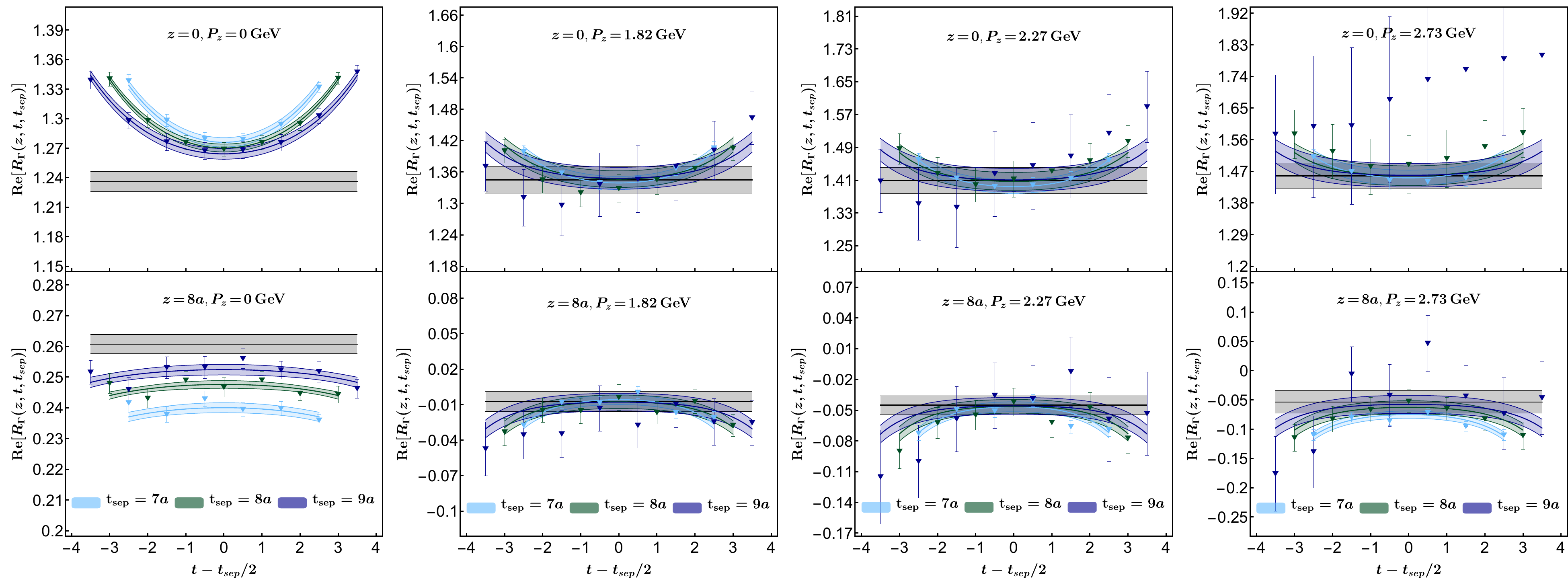}
\caption{{{Demonstration of fitting the ratio $\mathrm{Re}[R_{\Gamma}(z,t_\mathrm{sep},t)]$ to obtain the bare ground-state nucleon matrix element. Here we present the results of $z=0$ and $z=8a$ for momentum $P_z=\{0,1.82,2.27,2.73\}~\mathrm{GeV}$ used in H102.
}}}\label{fig:fitcmpr}
\end{center}
\end{figure*} 

\begin{figure*}[htbp]
\begin{center}
\includegraphics[width=1\textwidth]{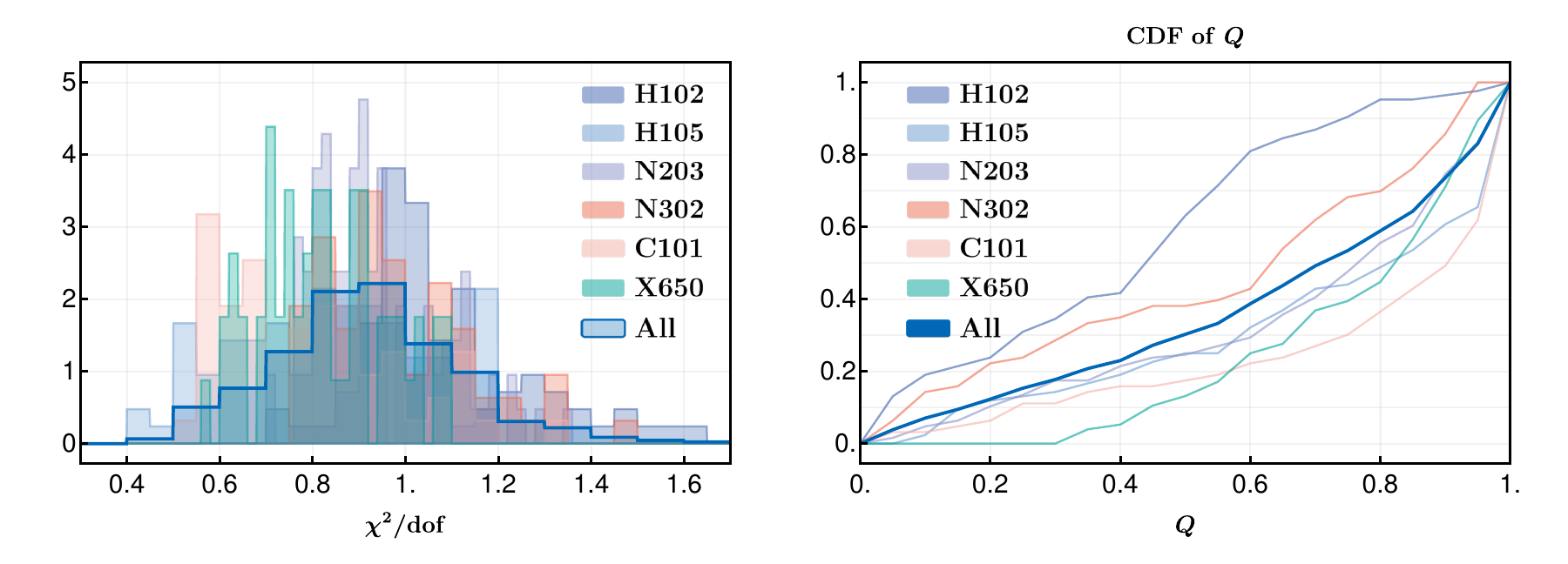}
\caption{{The histogram distributions {(normalized to 1)} of $\chi^2/\mathrm{d.o.f}$ and cumulative distribution function of $Q$-value collected from two-states fit, including all momenta and link length used in this work.}}\label{fig:fitchi}
\end{center}
\end{figure*}

In Table~\ref{Tab:tsepsetup}, we list the information on the nucleon momenta, $t_{\rm sep}$ and the number of measurements per configuration used on each ensemble. For the X650 ensemble, in RUN I (1000 configurations) we used one source and in RUN II (500) we used two, separated in space direction.

\begin{table}[htbp]
\centering
\renewcommand{\arraystretch}{1.4}
\resizebox{1.\columnwidth}{!}{
\begin{tabular}{cccccccccc} 
Ensemble  ~~&$P_z$[GeV]  ~~&$t_{\mathrm{sep}}/a$ ~~&$N_{\text{meas.}}/N_{\text{conf.}}$\\
\hline
\hline
\arrayrulecolor{gray!50}
X650    ~~& \{0, 1.84, 2.37, 2.63\}    ~~&\{5,7,9\}             ~~& {1\; (RUN I\phantom{I})}      \\
   ~~&  ~~&        ~~& {2\;(RUN II)}   \\\hline
H102    ~~& \{0, 1.82, 2.27, 2.73\}    ~~&\{7,8,9\}             ~~& 2      \\\hline
H105    ~~& \{0, 1.82\}    ~~&\{7,8,9\}             ~~& 2       \\\hline
C101   ~~& \{0, 1.82\}     ~~& \{6,7,8,9\}            ~~& 2   \\\hline
N203   ~~& \{0, 1.62, 2.02, 2.43, 2.83\}  ~~& $\begin{array}{c} \{10,11,12\\13,\\14,15\}\end{array}$  ~~& $\begin{array}{c}4\\8\\16\end{array}$     \\\hline
N302   ~~& \{0, 2.09, 2.62\}    ~~& $\begin{array}{c} \{10,12\\14,\\16,18\}\end{array}$    ~~& $\begin{array}{c}4\\8\\16\end{array}$ \\
\arrayrulecolor{black}
\hline
\end{tabular}}
 \caption{Information on nucleon momenta, $t_{\rm sep}$ and the number of measurements per configuration used in each ensemble. For N203 and N302, we increase the number of measurements for large $t_{\mathrm{sep}}$ to boost the signal. }
 \label{Tab:tsepsetup}
\end{table}

To extract the ground state matrix elements from lattice calculated two- and three-point functions, we use the following fitting form
\begin{align}
&C^\text{2pt}(t_\text{sep}) \approx c_4e^{-E_0 t_\mathrm{sep}}(1+c_5 e^{-\Delta E t_\mathrm{sep}})\,,\nn\\
&R_{\Gamma}(z,t,t_\text{sep}) \equiv \frac{C^\text{3pt}(z,t_\text{sep},t)}{C^\text{2pt}(t_\text{sep})},\nn\\
&\approx\frac{c_0(z)\!+\! c_1(z)\left[e^{-\Delta E (t_\mathrm{sep}-t)}\!+\!e^{-\Delta E t}\right]\!+\!c_3(z)e^{-\Delta E t_\mathrm{sep}}}{1+c_5 e^{-\Delta E t_{\mathrm{sep}}}},
\end{align}
where $c_0$ is the desired ground state matrix element $\widetilde{h}(z,P_z,1/a)$. We have neglected states beyond the ground state and first-excited state in our analyses. To justify this, we perform fits with different sets of source-sink separations $t_{\mathrm{sep}}$. Different $t_{\mathrm{sep}}$ analyses are consistent within statistical errors, which suggests that the excited-state contamination is well under control. 

In Fig.~\ref{fig:fitcmpr}, we present the data and results of fits to $R_{\Gamma}(z,t_\mathrm{sep},t)$ for H102. Here we show the results of $z=0$ and $z=8a$ for different momenta used in our analysis. As shown in {the figure}, we find good agreement between the data and fits. 

In Fig.~\ref{fig:fitchi} we present the histogram distribution (normalized to 1) of $\chi^2/\mathrm{d.o.f}$ (d.o.f represents degrees of freedom) and the cumulative distribution function (CDF) of the $Q$ values for our two-state fits. 
We provide the distributions for each ensemble, the distributions of combing all ensembles are also plotted. Most of the $\chi^2/\mathrm{d.o.f}$ are centered around 1 and only $3.8\%$ of the fits have $Q<0.05$ which illustrates the high quality of the fits.



\subsubsection{Large $\lambda$ extrapolation}

\begin{figure*}[htbp]
\begin{center}
\includegraphics[width=.8\textwidth]{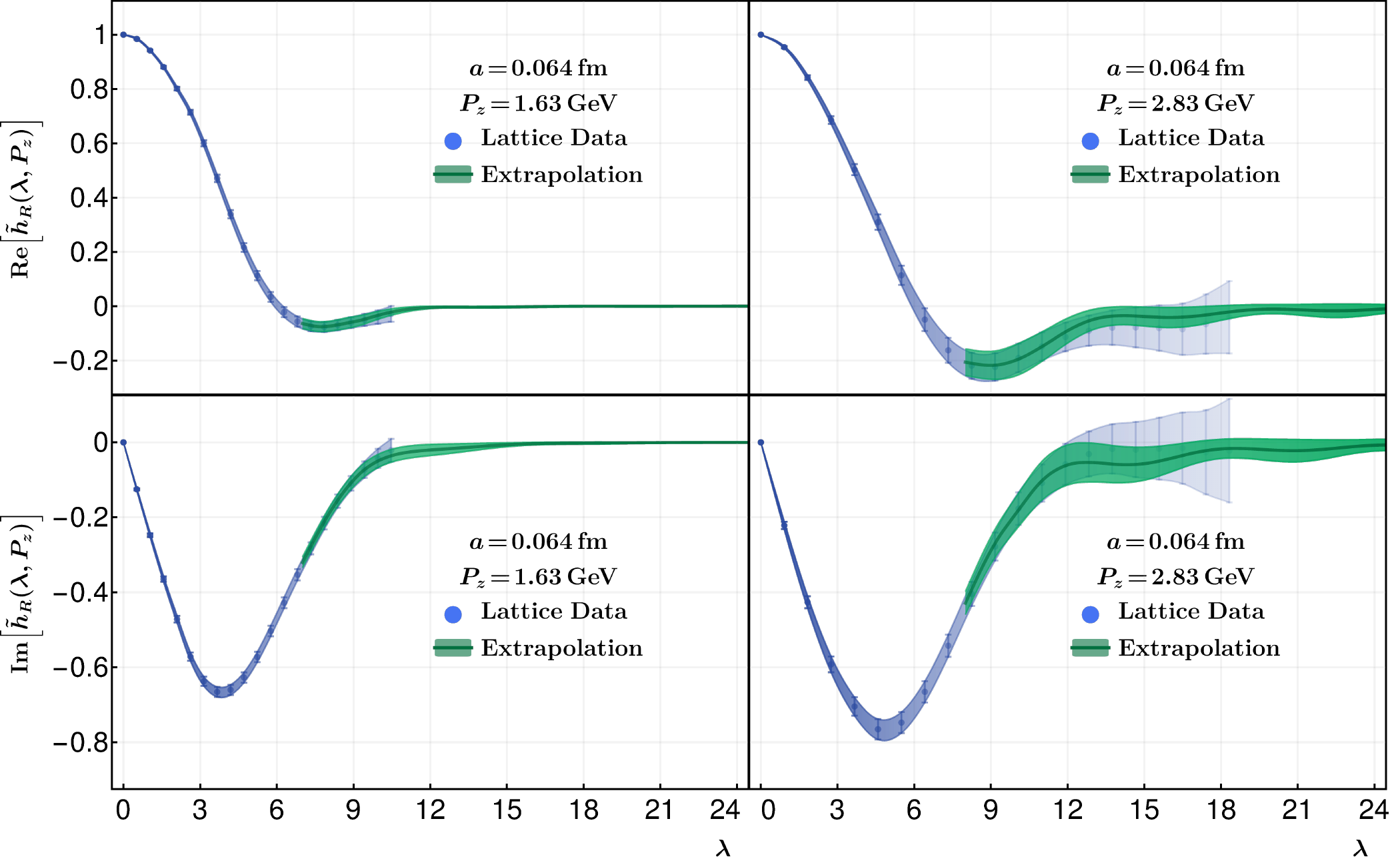}
\caption{The renormalized matrix elements $\tilde{h}_R(\lambda,P_z)$ and their extrapolation. The extrapolation reproduces the lattice data in the moderate $\lambda$ region, and yields smooth correlations with controllable uncertainties in the large $\lambda$ region.}\label{fig:extrplt}
\end{center}
\end{figure*}

{For the large $\lambda$ region, we use the physics-based exponential form Eq.~(6\vphantom{\ref{eq:extrap}}) of the main text to supplement the renormalized quasi-LF correlation calculated on the lattice. To this end, a reasonable large $\lambda$ region {{($\lambda \geq \lambda_L$)}} needs to be chosen to determine the extrapolation parameters. {{$\lambda_L$ is regarded as a truncation value.}} In Fig.~\ref{fig:extrplt}, we show the extrapolation of the renormalized matrix elements on N203 with one small momentum ($P_z=1.63\,\mathrm{GeV}$) and one large momentum ($P_z=2.83\,\mathrm{GeV}$) as example. {{We choose $\lambda_L=7$ for the extrapolation, and vary it down to $\lambda_L=4$}} to estimate the systematic error from extrapolation. As can be seen from the figure, the fitting results agree with the original lattice data in the moderate $\lambda$ region, and give smooth correlations at large $\lambda$. By doing this, we avoid unphysical oscillations in the momentum distribution obtained by a brute-force Fourier transform, with the price that the  behavior in the endpoint region (conjugate to large-$\lambda$ region) is altered. However, this is the region that cannot be reliably predicted by LaMET anyway.}

\subsubsection{Dependence of momentum space distributions on $m_\pi$ and $P_z$ and combined extrapolation}

\begin{figure}[htbp]
\includegraphics[width=.45\textwidth]{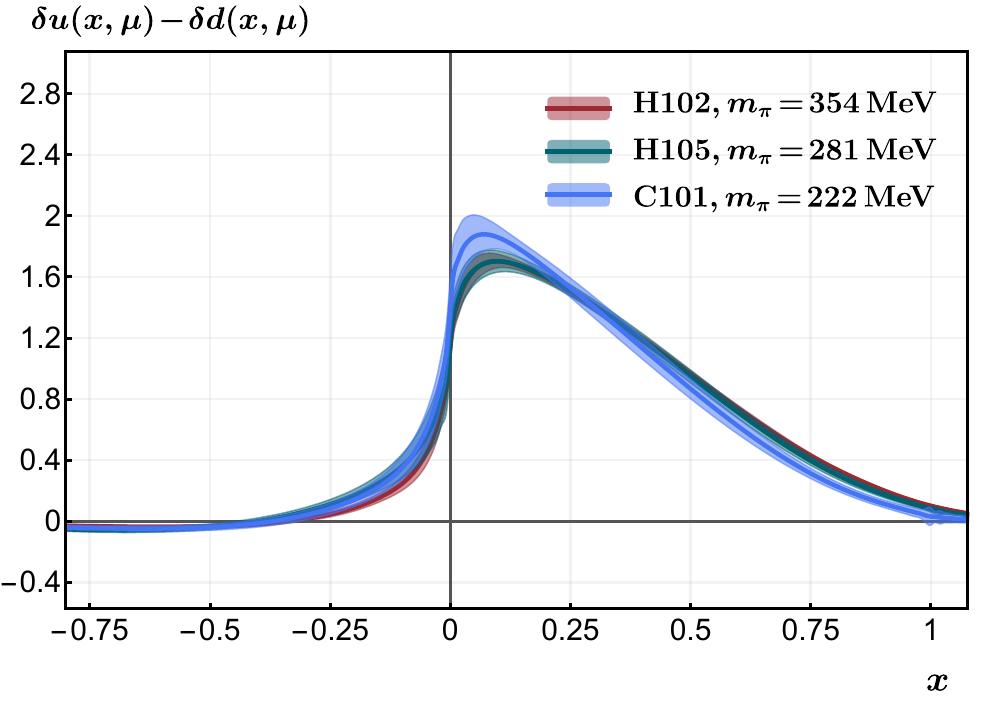}
\caption{The pion mass dependence of the extracted PDFs on different ensembles with the same lattice spacing $a=0.085$~fm and nucleon momentum $P_z=1.82$~GeV.}
\label{fig:piondepxspace}
\end{figure}

\begin{figure}[htbp]
\includegraphics[width=.46\textwidth]{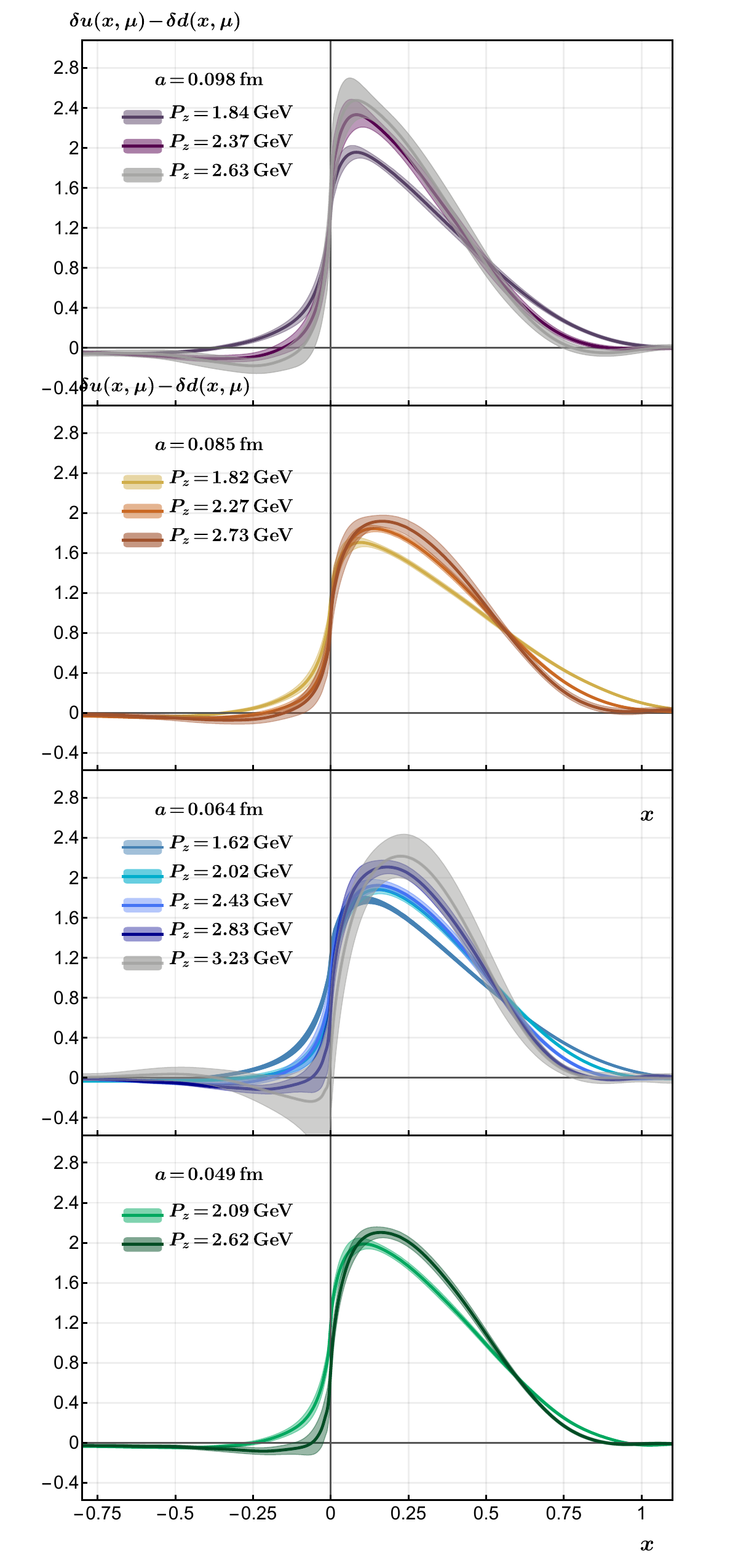}
\caption{The momentum dependence of the extracted PDF on X650, H102, N203 and N302 with $m_\pi \approx 350$ MeV. Grey bands are excluded due to large uncertainties in the combined extrapolation.  }
\label{fig:pzdepxspace}
\end{figure}

For our lattice setup, it is difficult to have the same momenta for ensembles of different lattice spacings. Thus, the momentum dependence and $a$-dependence cannot be easily disentangled. They are removed by a combined extrapolation to be discussed later. 
In Fig.~\ref{fig:piondepxspace}, we show the pion mass dependence of the  PDFs extracted from data at the same lattice spacing and nucleon momentum. The results contain only statistical errors.
As can be seen from the figure, the extracted PDFs with $m_\pi=\{354,281,222\}$ MeV almost coincide with each other within errors, indicating a very mild pion mass dependence.
%
%
In Fig.~\ref{fig:pzdepxspace}, we show the momentum dependence of the extracted PDFs from ensembles of roughly the same pion mass, which is stronger than the pion mass dependence. 

\begin{figure*}[htbp]
\begin{center}
\includegraphics[width=0.9\textwidth]{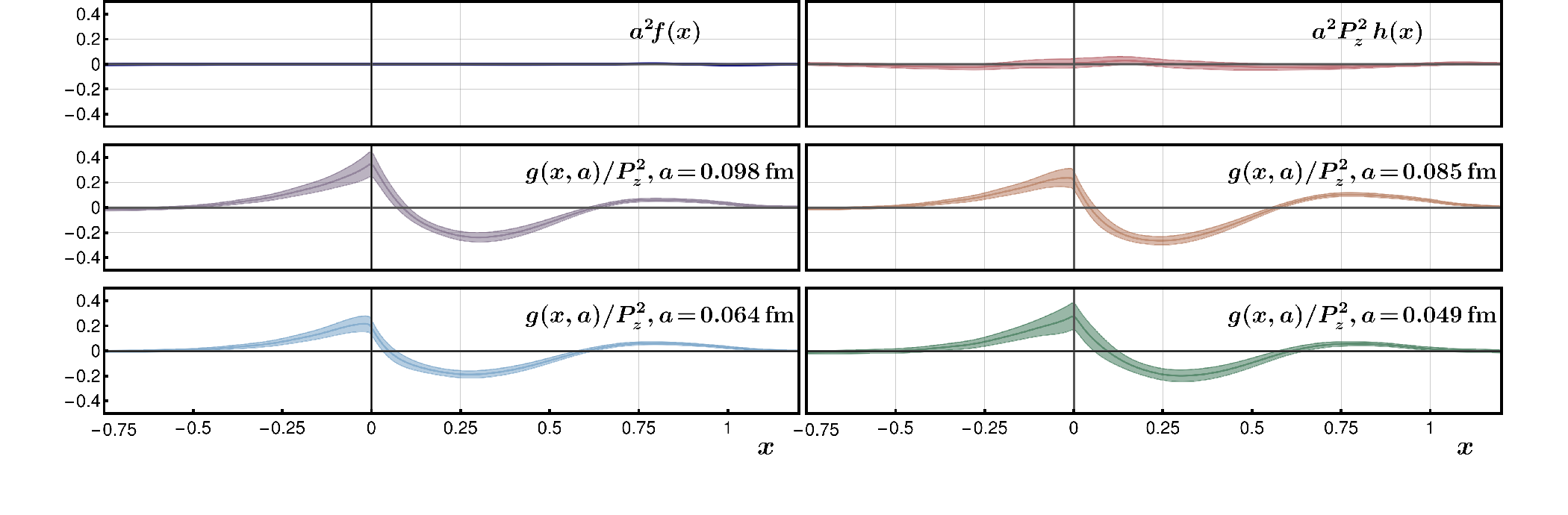}
\caption{Estimate of the size of discretization effects (first row) and {higher-twist corrections (second and third rows)}.}
\label{fig:parafig}
\end{center}
\end{figure*} 
To obtain the transversity PDF in the continuum, physical pion mass and infinite momentum limit, we do a combined extrapolation using the following form,
    \begin{align}\label{eq:combinedfitsupp}
    	\delta q\left(x,P_z,a, m_\pi\right)&=\frac{1-g' m_\pi^2\,\ln (m_\pi^2/\mu_0^2)+m_\pi^2 k(x)}{1-g' m_\pi^2\,\ln (m_\pi^2/\mu_0^2)}\nn\\
    	&\hspace{-4em}\times\left[\delta q_0(x)+a^2f(x)+a^2P_z^2 h(x)+\frac{g(x,a)}{P_z^2}\right],
    \end{align}
where $\delta q\left(x,P_z,a, m_\pi\right)$ denotes the extracted transversity PDFs with different $a$, $P_z$ and $m_\pi$ depicted in Figs.~\ref{fig:piondepxspace} and \ref{fig:pzdepxspace}. In addition to the $m_\pi^2 k(x)$ term for the CLS setup~\cite{RQCD:2019hps}, we also include the chiral logarithm $g' m_\pi^2 \ln (m_\pi^2/\mu_0^2)$ in the pion mass extrapolation with $\mu_0=1$ GeV, $g'=-(4g_A^2+1)/[2(4\pi f_\pi)^2]$ and $g_A$ being the axial charge of the nucleon~\cite{Chen:2001eg}. 
The denominator of Eq.~(\ref{eq:combinedfitsupp}) follows from the fact that our extracted PDF is normalized by the tensor charge $g_T$ whose chiral extrapolation is given in Ref.~\cite{Chen:2001eg}. The $a^2 f(x)$ and $a^2 P_z^2 h(x)$ denote discretization effects. The last term in the square bracket, $g(x,a)/P_z^2$, characterizes the dependence of the power correction on $P_z$, where we also keep an explicit $a$ dependence in the numerator. Finally, the desired transversity PDF is given by
\begin{align}
\delta q\left(x\right)&\!=\!\frac{1\!-\!g' m_{\pi,\rm phys}^2\ln (m_{\pi,\rm phys}^2/\mu_0^2)\!+\!m_{\pi,\rm phys}^2 k(x)}{1\!-\!g' m_{\pi,\rm phys}^2\,\ln (m_{\pi,\rm phys}^2/\mu_0^2)}\delta q_0(x).
\end{align}


\begin{figure*}[htbp]
\begin{center}
\begin{minipage}{0.45\linewidth}
\includegraphics[width=0.85\linewidth]{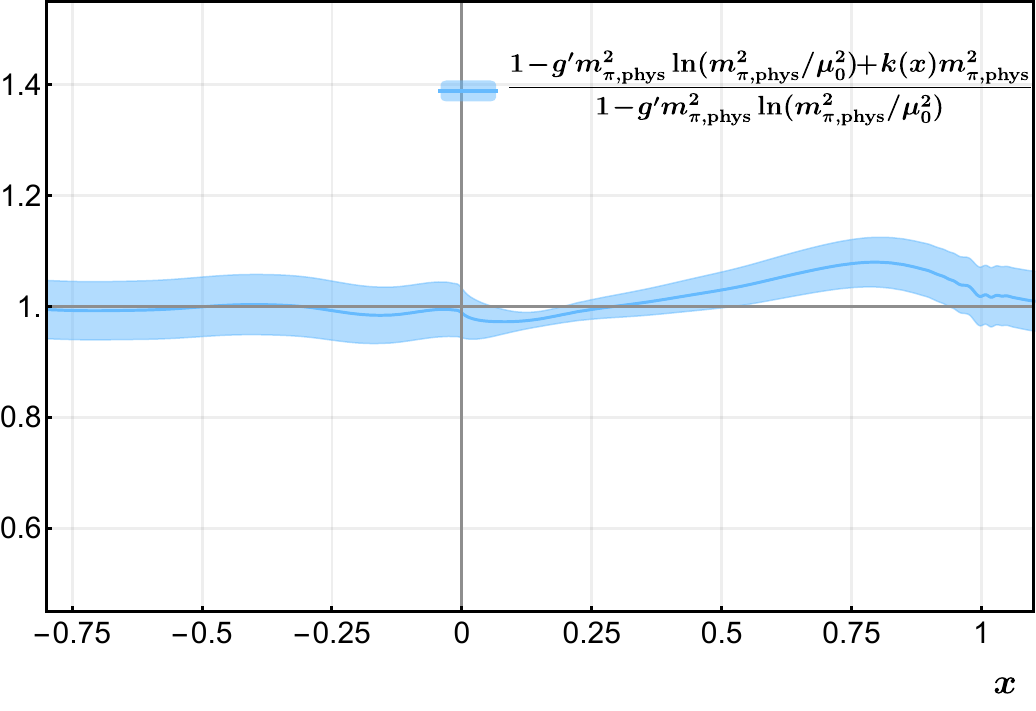}
\caption{Estimation of the size of $m_\pi$ dependence. The blue band is fixed by taking $g_A=1$, $f_\pi=93$ MeV and $\mu=2$ GeV in $g'$.\\ \\ \\} 
\label{fig:parafigfx}
\end{minipage}\quad
\begin{minipage}{0.45\linewidth}
\includegraphics[width=0.9\linewidth]{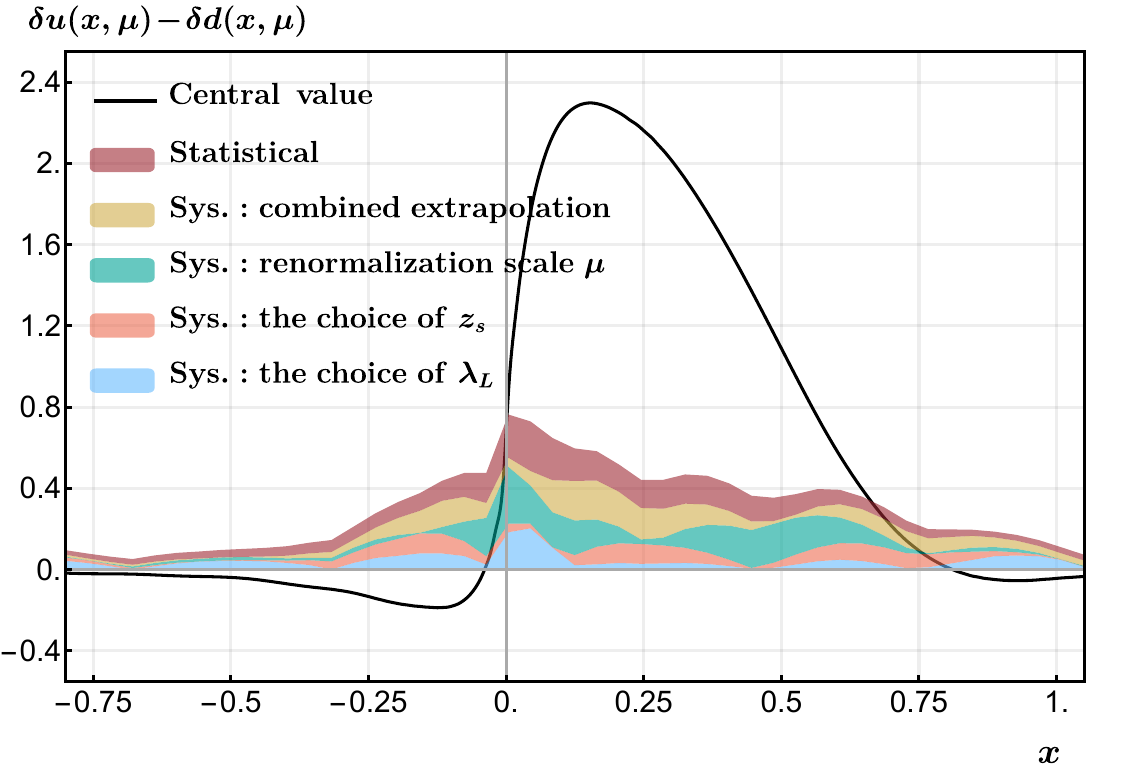}
\caption{Estimation of statistical and systematic uncertainties. The width of the (non-overlapping) coloured bands denotes the size of each uncertainty (these uncertainties are added in quadrature to obtain the full systematic uncertainty), and the black line is the extrapolated central value obtained at $z_s=0.3$~fm, $\mu=2$ GeV and $\lambda_L=7$. }
\label{fig:syserr}
\end{minipage}
\end{center}
\end{figure*}

In Fig.~\ref{fig:parafig} we show the size of discretization effects and higher-twist corrections. The $a^2 f(x)$ term is nearly zero, because it has largely been taken into account in the self-renormalization Eq.~(\ref{eq:lnbrMEFit}). The $a^2P_z^2\,h(x)$ term also turns out to be close to zero. 
In contrast, the $g(x,a)/P_z^2$ term is much larger and is the dominant uncertainty from the extrapolation.

The $m_\pi$ dependence reflects itself in the ratio in front of the square bracket in Eq.~({\ref{eq:combinedfitsupp}}). We plot this ratio for different $x$ in Fig.~\ref{fig:parafigfx}. As can be seen from the figure, the ratio is very close to one in most of the $x$ region, except in the neighborhood of $x\approx0.75$ where it is slightly larger, indicating a mild pion mass dependence of our results. 
One of the reasons is that the chiral logarithm in the numerator of the ratio is largely canceled by the same logarithm in the denominator.

\subsection{Final result}

\subsubsection{Systematic uncertainties}

 The central value and the statistical error of the final result is obtained from the PDF results extracted from all configurations at $z_s=0.3$~fm, $\mu=2$~GeV and {$\lambda_L=7$}, where the statistical error propagates to the final result according to the combined extrapolation formula. The error band includes four systematic uncertainties apart from the statistical uncertainty:
 
a. Combined extrapolation ($a \rightarrow 0$, $P_z \rightarrow \infty$, $m_\pi \rightarrow 135$~MeV). In our analysis, we have performed a combined extrapolation in $a$, $P_z$ and $m_\pi$, as discussed in the previous section. To account for the systematic uncertainty due to this extrapolation, we take the difference between the central value of the combined extrapolation result and the PDF result extracted at the smallest lattice spacing $a=0.049$~fm with large pion mass $m_\pi=348$~MeV and momentum $P_z=2.62$~GeV as an estimate. 
\\
b. Renormalization scale dependence. 
The systematic uncertainty due to the choice of renormalization scale is estimated by varying $\mu$ from 2 to 3 GeV. Specifically, we repeat the calculation of renormalization, Fourier transform, matching and combined extrapolation at $\mu=3$ GeV, then take the difference between this extrapolated result and the central value at $\mu=2$ GeV. \\
c. The choice of $z_s$ in the hybrid scheme. The error from varying $z_s$ is obtained by taking the difference between the results at $z_s=0.3$ fm and $z_s=0.18$ fm. Actually, for $z_s$ in the region where the self-renormalization and the ratio scheme results coincide, the dependence on $z_s$ is very weak. A comparison of this uncertainty with other systematic uncertainties is given in Fig.~\ref{fig:syserr}, where choosing different $z_s$ only leads to a small uncertainty. \\
d. The choice of $\lambda_L$. We have chosen a different extrapolation region $\lambda_L=4$, and taken the difference in the result as an estimate of the systematic uncertainty due to $\lambda$ extrapolation. As can be seen from Fig.~\ref{fig:syserr}, the small $x$ region is more sensitive to the $\lambda$ extrapolation, as expected from theory. \\

In Fig.~\ref{fig:syserr}, we show the central value of the transversity PDF and all uncertainties, where the width of each colored band denotes the size of the corresponding uncertainty. 


\begin{figure*}[htbp]
\begin{center}
\includegraphics[width=0.95\textwidth]{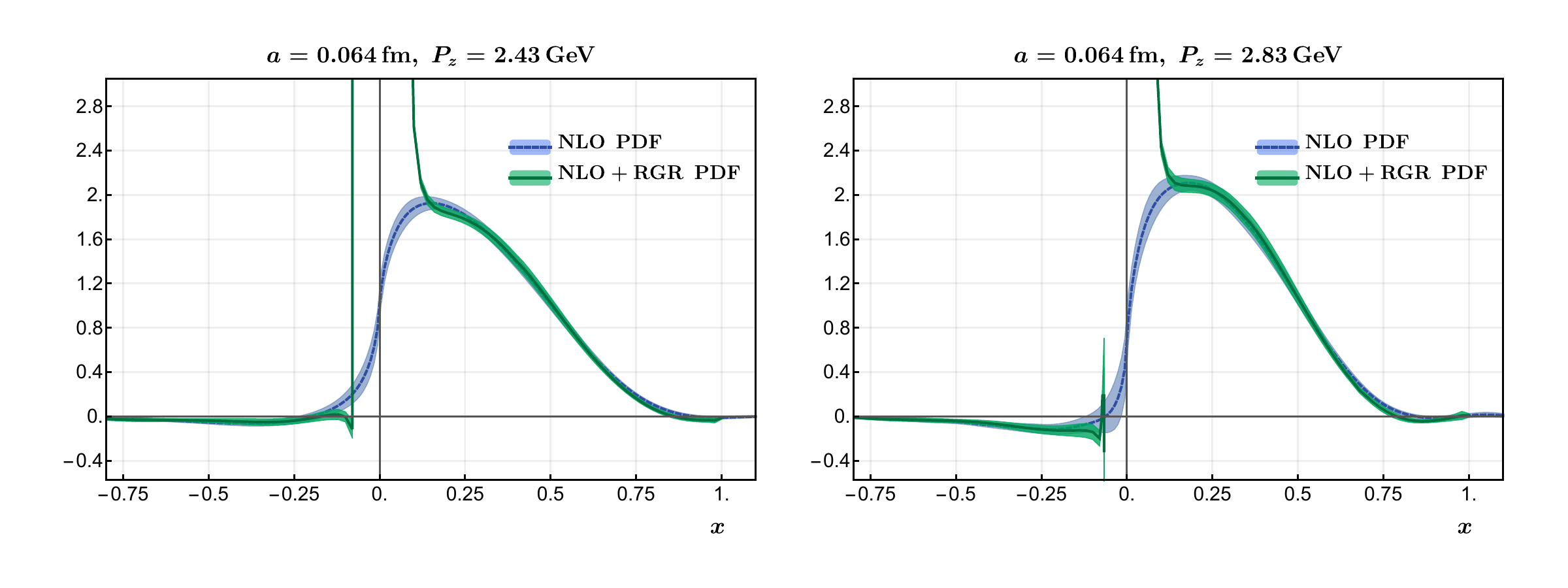}
\caption{Comparison between results incorporating the RGR and fixed-order matching. We plot the cases $a=0.064$~fm and $P_z=\{2.43,2.83\}~\mathrm{GeV}$ for demonstration.}
\label{fig:N203RGR}
\end{center}
\end{figure*} 

\subsubsection{Renormalization group resummation for the end-point region}
In Ref.~\cite{Su:2022fiu}, it was pointed out that in the small $x$ region, large logarithms could appear and a renormalization group resummation (RGR) is required. Taking N203 as an example, we present the light-cone transversity PDF in Fig.~\ref{fig:N203RGR} by applying the RGR to our analysis. As can be seen from the figure, the RGR-improved transversity PDF blows up in the small $x$ region ($|x|\lesssim 0.1$), clearly indicating that LaMET is unreliable in this region. 
Note that the reliable regions estimated from the RGR analysis is consistent with the naive {estimate from the criterion that the higher-twist terms $\Lambda_\mathrm{QCD}^2/(xP_z)^2$ are of $\mathcal{O}(1)$}. On the other hand, in the region $x\rightarrow1$, the relative error is actually large, but it is hard to see in the figure as the central value is very close to zero.


\bibliographystyle{apsrev}
\bibliography{zsupp}

\clearpage

\end{document}